\documentclass[12pt]{article}
\pdfoutput=1
\usepackage[a4paper]{geometry}
\usepackage{jheppub, amsmath,amssymb,amsfonts,amsxtra, mathrsfs, makeidx,graphics,graphicx,amsthm,epsfig, bm,longtable,float, color,tikz,mathtools,xfrac,footnote,rotating, lscape, makecell, environ,mathtools, empheq}

\usepackage[T1]{fontenc} 

\usepackage{subfig}
\usepackage{multirow}
\usepackage{adjustbox}

\usepackage{amsthm}
\usepackage{hyperref}

\usepackage{amsmath}
\usepackage{amsfonts}
\usepackage{commath}
\usepackage[english]{babel}
\usepackage{setspace}
\usepackage{youngtab}

\usepackage{epsfig,amssymb} 
\usepackage{amsmath}

\usepackage{amscd,color}
\usepackage{amsmath,graphicx}
\usepackage{verbatim,booktabs}

\usepackage{latexsym}
\usepackage{amsmath,amsfonts,amssymb,amsthm}
\usepackage{amsmath,amsthm}

\newcommand{\nn}{n}

\usepackage{tikz}
\usetikzlibrary{tikzmark}
\usetikzlibrary{math} 
\usetikzlibrary{matrix}
\usetikzlibrary{calc}
\usetikzlibrary{positioning}
\definecolor{azzurro}{RGB}{0,118,186}
\definecolor{rosso}{RGB}{255,0,0}

\usepackage{array}

\definecolor{mandarancio}{RGB}{255,147,0}
\definecolor{azzurro}{RGB}{0,118,186}
\definecolor{terrabruciatadiconcorezzo}{RGB}{181,23,0}

\def\centerarc[#1](#2)(#3:#4:#5)
{ \draw[#1] ($(#2)+({#5*cos(#3)},{#5*sin(#3)})$) arc (#3:#4:#5) }

\newcommand{\munutable}[6] 
{
\begin{equation}
\begin{array}{|c|l|l|}
\hline r=#2 & \multicolumn{2}{c|}{#1}\\
\hline \nu \backslash \mu & \multicolumn{1}{c|}{1} & \multicolumn{1}{c|}{-1} \\
\hline 1 &\begin{array}{l} #3 \end{array} &\begin{array}{l} #4 \end{array} \\
\hline-1 & \begin{array}{l} #5 \end{array} & \begin{array}{l} #6 \end{array} \\
\hline
\end{array}
\end{equation}
}

\title{
\begin{center}
Symplectic gauge group on the Lens Space
\end{center}
}

\author[a]{Antonio 
Amariti}
\author[a,b]{and Simone 
 Rota}
\affiliation[a]{INFN, Sezione di Milano, Via Celoria 16, I-20133 Milano, Italy}

\affiliation[b]{Dipartimento di Fisica, Universit\`a degli Studi di Milano, Via Celoria 16, I-20133 Milano, Italy}
\emailAdd{antonio.amariti@mi.infn.it,  simone.rota@mi.infn.it}

\abstract{
We compute the Lens space index for 4d supersymmetric gauge theories involving symplectic gauge groups.
This index can distinguish between different gauge groups from a given algebra and
it matches across theories related by supersymmetric dualities.
We provide explicit calculations for $\mathcal{N}=4$ SYM and for classes of $\mathcal{N}=2$ and $\mathcal{N}=1$ Lagrangian quivers related by S-duality. In these cases the index matches across the S-dual phases, 
while models in different S-duality orbits have a different Lens index.
We provide analogous computations for a 4d $\mathcal{N}=1$ toric quiver gauge theory corresponding to a $\mathbb{Z}_7$ orbifold of $\mathbb{C}^3$. This $SU(n)^7$ gauge theory becomes interesting in the case of $n=2$ because it is conformally dual 
to other two models, with symplectic and unitary gauge groups, bifundamentals and antisymmetric tensors.
We explicitly check this triality at the level of the Lens space index. 
}

\begin{document}

\maketitle

\section{Introduction}

Higher forms play a crucial role in the modern formulation of symmetries in QFT, because they 
involve extended object and  constrain their charges \cite{Gaiotto:2014kfa}.
For example the extended objects charged under a 1-form symmetry are loop operators.
The constraints on the charge spectrum of such lines, i.e. Wilson and 't Hooft lines, is reflected in the choice of 
the gauge group from a given gauge algebra in a gauge theory \cite{Gaiotto:2010be,Aharony:2013hda}.   
Fixing the global properties of a gauge theory can have important consequences, for example it fixes the periodicity of the
theta angle (see for example \cite{Tong:2017oea} for the phenomenological implication in the SM).

Another more formal consequence of having different gauge groups for a given gauge algebra is that this difference 
can be observed on the partition function  computed in curved space on a spin manifold \cite{Aharony:2013hda,Razamat:2013opa}. In general partition functions on curved space  are complicated quantities, but the difficulty of such problem is highly simplified for some manifolds in supersymmetric gauge theories, thanks to the help of localization \cite{Pestun:2007rz}. 

In general such partition functions can still fail in distinguishing among different global structures because of the presence of further symmetries and dualities relating theories with different gauge groups. This is for example the case of S-duality in $\mathcal{N}=4$ SYM.
In this case many of the choices of the global structure are related to each other by the action of the  S-duality group
on the spectrum of  charges of the line operators. There are nevertheless, depending on the choices of the gauge algebras and the ranks, cases where multiple orbits of the S-duality group are present. The partition functions on the curved space can in principle distinguish such orbits.
This picture has been confirmed by explicit calculation from the Lens space index in \cite{Razamat:2013opa}. This index was originally defined in \cite{Benini:2011nc}  and it corresponds to  the superconformal index computed 
on $L(r,1)\times S^1$, where  $L(r,1) \simeq S^3 / {\mathbb{Z}_r}$ is the three-dimensional Lens space and $S^1$ is the Euclidean time.

The first explicit calculations of the index on such space have been performed in  \cite{Razamat:2013opa}  for the case of $\mathcal{N}=4$ SU(n).  Furthermore Seiberg duality for $SO(n)$ gauge theories with vectors has been analyzed as well.
Other calculations of the Lens space index have been performed in \cite{Amariti:2019but} for the $\mathcal{N}=2$ $SU(n)$ quiver corresponding to the non-chiral orbifolds of $\mathbb{C}^3$ (see also \cite{Alday:2013rs,Razamat:2013jxa,Fluder:2017oxm} for an analysis for non lagrangian theories).
In all the cases the index has been shown to match for models connected by duality while it gave different results for different orbits of the S-duality group.

The 4d Lagrangian SCFTs zoology admits however many other possibile behaviors that have not yet be studied in terms of the
Lens space index and that require an investigation.
For example models with symplectic gauge groups have not been analyzed so far. This comprises the case of $\mathcal{N}=4$ where depending on the parity of the gauge rank we have different structure of the S-duality orbits, involving orthogonal groups as well \cite{Aharony:2013hda}.
Furthermore symplectic, orthogonal and unitary gauge groups are all involved in the examples of S-duality for $\mathcal{N}=2$ 
quivers originally found in \cite{Uranga:1998uj} that can be studied in terms of the Lens space index.
Such S-duality has been recently shown in \cite{Amariti:2021lhk} to persist when breaking $\mathcal{N}=2$ to $\mathcal{N}=1$.
In this paper we study the Lens space index for these models, showing that they match among the 
theories in the same S-duality orbit, while they differ for choices of the gauge group in a different orbit.

We conclude our analysis by studying a triality found in \cite{Razamat:2020pra} that relates three models with either unitary or 
symplectic gauge groups. It was pointed out in \cite{Razamat:2020pra} that there are in these cases different choices of the gauge group for each phase and we observe that 
all these choices give rise to the same Lens space index. On one hand this corroborates the validity of the claim about the triality among these models. 
On the other hand we explain the absence of multiple orbits by discussing some general expectations from the holographic dual description of one of these three phases in Type IIB string theory.

\section{Review}
Extended operators, such as Wilson and t'Hooft lines, play an important role in the study of Quantum Field Theories. When the spacetime manifold is $\mathbb{R}^4$ the extended operators of the theory do not affect the correlation function of local operators. Nevertheless, two theories that only differ by their extended operators are still distinguished by the correlation functions that involve the extended operators themselves. When the spacetime manifold is non-trivial the extended operators can have a wider impact on the physics. Extended operators can be wrapped on non-trivial cycles of the spacetime providing different backgrounds for the local physics. Furthermore when the theory is compactified to a lower dimension the presence of extended operators can change the spectrum of local operators on the lower-dimensional theory. For example when the spacetime is $\mathbb{R}^3 \times S^1$ a Wilson line wrapped around $S^1$ becomes a local operator in the effective 3-dimensional theory on $\mathbb{R}^3$.\\

\subsection{Line operators in gauge theories}
\begin{figure}
	\centering
	\includegraphics[scale=0.5]{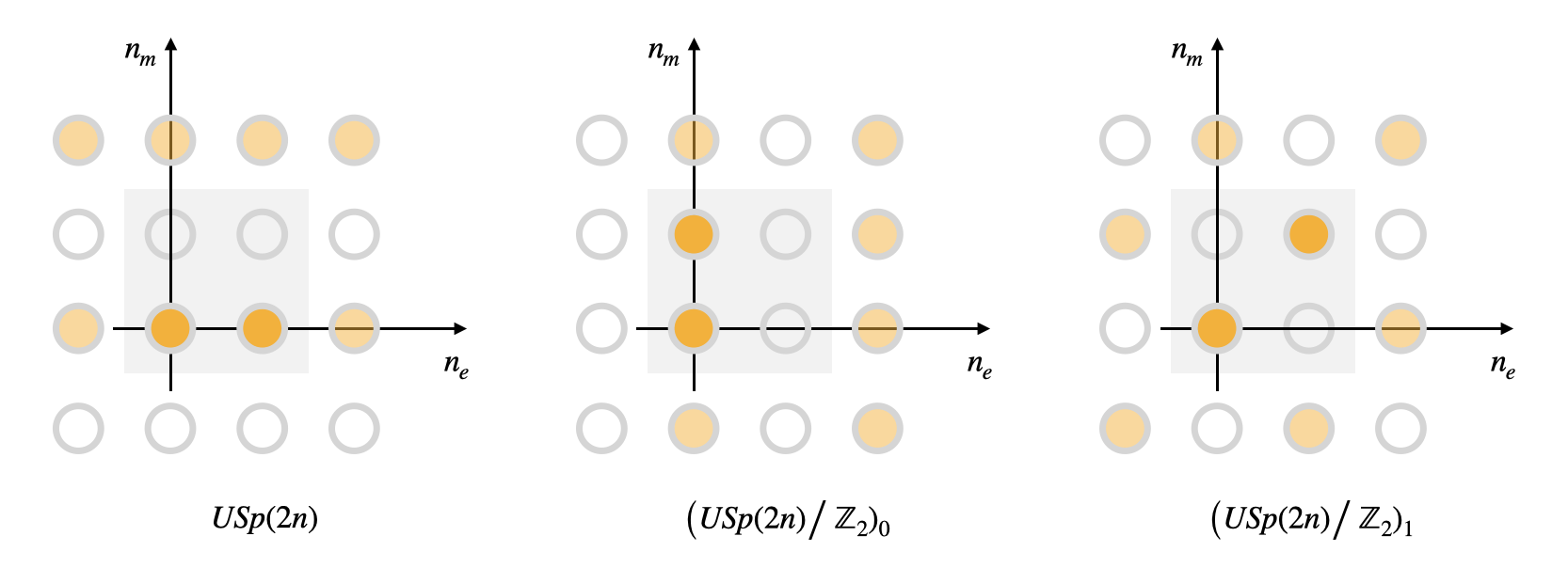}\
	\caption{The three possible choices of line charges for gauge theories with gauge algebra $\mathfrak{usp}(2n)$. The orange dots represent the charges of the line operators that are included in the corresponding theory.}
	\label{fig:usp_line_lattice}
\end{figure}

In gauge theories the spectrum of extended operators is closely related to the global structure of the gauge group. The spectrum and correlation functions of local operators only depend on the gauge algebra $\mathfrak{g}$ associated to the gauge group $G$ while the spectrum of lines depends on the gauge group $G$ and on discrete theta-like parameters.
In this paper we will only consider compact gauge groups, therefore we have $G = \tilde{G} / H$ where $\tilde{G}$ is the compact simply connected group with associated Lie algebra $\mathfrak{g}$ and $H \in \mathcal{Z}(\tilde{G})$ is a subgroup of the center of $\tilde{G}$.
The lines can be organized by their electric and magnetic charges $(n_e, n_m) \in \mathcal{Z}(\tilde{G})\times \mathcal{Z}(\tilde{G})$. We always have Wilson lines in every representation of $G$ that belong to the classes $(n_e, 0)$ with $n_e$ invariant under the action of $H$. These are completely determined by the choice of gauge group $G$. In addition to the Wilson lines the theory includes t'Hooft and dyonic lines. Any two lines of the theory must satisfy a Dirac pairing condition, for example when $\mathcal{Z}(G) = \mathbb{Z}_k$ the condition reads:
\begin{equation} \label{eq:Dirac_pairing}
	n_e n_m' - n_e' n_m = 0 \, \text{mod} \, k
\end{equation}
The spectrum of lines is determined by a complete and maximal set of charges $(n_e , n_m)$ satisfying \eqref{eq:Dirac_pairing}. It turns out that given a choice of gauge group $G$ there still can be different choices for the spectrum of lines. These choices are associated to discrete theta-like parameters that can be introduced in the theory. For example when $\mathfrak{g}=\mathfrak{usp}(2n)$ there are three possible choices for the line spectrum, they are depicted in Figure \ref{fig:usp_line_lattice}.\\

\subsection{Lens space index}
The Lens space index is a powerful tool for studying the global structure of supersymmetric gauge theories \cite{Razamat:2013opa, Amariti:2019but, Schweigert:1996tg, Kels:2017toi,Kels:2017vbc, Fluder:2017oxm} . Unlike the supersymmetric index on $S^3 \times S^1$, the Lens space index is sensible to the extended $(1+1)$-dimensional operators of the theory (i.e. Wilson, t'Hooft and dyonic lines) and can be able to distinguish between theories with the same gauge algebra and matter content but with different gauge group.
The index is an RG invariant and is expected to match between dual theories as well as being stable under exactly marginal deformations. The Lens space index of a theory can be computed as a supersymmetric partition function on the (Euclidean) spacetime manifold $L(r,1)\times S^1$. Here $L(r,1) \simeq S^3 / {\mathbb{Z}_r}$ is the three-dimensional Lens space and $S^1$ is the Euclidean time. The integer $r$ parametrizes different spacetime manifolds, explicitly:
\begin{equation}
L(r,1) = \lbrace x_1, x_2 \in \mathbb{C} \, s.t.\, |x_1|^2 + |x_2|^2 =1 \rbrace_{\sim_r}
\end{equation}
where we identify elements related by $\sim_r$:
\begin{equation}
	(x_1, x_2) \sim_r (e^{\frac{2\pi i }{r}} x_1,e^{- \frac{2\pi i }{r}} x_2)
\end{equation}

The fundamental group of the Lens space is:
\begin{equation}
	\pi_1 \left( L(r,1)\times S^1 \right) = \mathbb{Z}\times \mathbb{Z}_r
\end{equation}
therefore there are two non-contractible 1-cycles: the Euclidean time cycle wrapping $S^1$ and a 1-cycle $\gamma$ in $L(r,1)$ such that $\gamma^r$ is contractible. When a gauge theory is placed on this spacetime one has to sum over all possible gauge bundles. In particular the allowed flat connections for a gauge group $G$ are determined by the holonomies of the gauge field around the non-contractible 1-cycles. We call $g$ and $h$ the holonomies around the cycle $\gamma$ and the Euclidean time cycle respectively. Two pairs of holonomies $(g,h)$ and $(g',h')$ are gauge equivalent if $g$ and $h$ can be simultaneously conjugated to $g'$ and $h'$ by an element of the group $G$. The sum over flat connection must be performed modulo gauge equivalence.

The choice of holonomies is a group homeomorphism $G \times G \to \pi_1 \left( L(r,1)\times S^1 \right)$ therefore $g$ and $h$ satisfy the following identities:
\begin{equation} \label{eq:commutator_g_h}
	g \cdot h \cdot g^{-1} \cdot h^{-1} = 1
\end{equation}

\begin{equation} \label{eq:gr_eq1}
	g^r = 1
\end{equation}

If $G$ is simply connected eq. \eqref{eq:commutator_g_h} implies that $g$ and $h$ can be simultaneously conjugated to a maximal torus of $G$. The same is not true if $G$ is not simply connected. The problem of commuting pairs\footnotemark has been studied for example in \cite{Borel:1999bx, Schweigert:1996tg, Witten:1997bs}, the solution for the groups associated to the algebra $\mathfrak{su}(n)$ and for $Spin(n)$, $SO(n)_\pm$ together with the constraint coming from \eqref{eq:gr_eq1} has been used in \cite{Razamat:2013opa} to compute the Lens space index of theories with the corresponding gauge group. The commuting pairs for symplectic groups has been studied e.g. in \cite{Borel:1999bx}, in \cite{Eager:2020rra} they have been used to compute the elliptic genus of two-dimensional theories with symplectic gauge group. In Appendix \ref{app:A} we summarize these results and show how to compute the Lens space index for $USp(2n)$ gauge theories.\\

\footnotetext{There is an analogous problem of commuting triples that is relevant for gauge theories compactified on the three-torus or other manifolds with three non-contractible 1-cycles. For further reading see \cite{Borel:1999bx}.}

A supersymmetric theory can be placed on the Lens space $L(r,1)\times S^1$ while preserving $\mathcal{N}=1$ supersymmetry. The corresponding partition function localizes on the flat connections of the gauge group and reduces to an effective matrix model. The infinite dimensional path integral reduces to a sum/integral over the holonomies $g$ and $h$. Each multiplet contributes to the integrand in the following way. Suppose that on the background of a specific choice of holonomies a field with R-charge\footnotemark $R$ acquire a phase $e^{\frac{2 \pi i m}{r}}$ when is rotated around the cycle $\gamma$ and a phase $u$ when is rotated around the euclidean time cycle. Then the contribution from a chiral multiplet is:
\footnotetext{In this paper we use notation where the R-charge $R$ of a chiral multiplet is the R-charge of the scalar field in the multiplet.}
\begin{equation}
	\mathcal{I}^{(R)}_\chi  (m,u) = \mathcal{I}_0 (m,u) \Gamma ((pq)^{\frac{R}{2}} q^{r-m} u; q^r,pq) 
					 \Gamma ((pq)^{\frac{R}{2}} p^{m} u; p^r,pq) 
\end{equation}
where $p$ and $q$ are fugacities associated to the spacetime symmetries $SU(2)_L \times SU(2)_R$ and 
\begin{equation}
\mathcal{I}_0(m, u)=\left((p q)^{\frac{1-R}{2}} u^{-1}\right)^{\frac{m(r-m)}{2 r}}\left(\frac{p}{q}\right)^{\frac{m(r-m)(r-2 m)}{12 r}} 
\end{equation}
is the contribution of the multiplet to the Casimir energy. The contribution of a vector multiplet is:
\begin{equation}
\mathcal{I}_V(m, u)=\frac{\mathcal{I}_0^V(m, u)}{\left(1-u^{-1}\right)^{\delta_{m, 0}} \Gamma\left(q^m u^{-1} ; q^r, p q\right) \Gamma\left(p^{r-m} u^{-1} ; p^r, p q\right)}
\end{equation}
where:
\begin{equation}
\mathcal{I}_0^V(m, u)=\left((p q)^{\frac{1}{2}} u^{-1}\right)^{-\frac{m(r-m)}{2 r}}\left(\frac{q}{p}\right)^{\frac{m(r-m)(r-2 m)}{12 r}} .
\end{equation}
The elliptic Gamma function $\Gamma(z;p,q)$ is defined by the infinite series:
\begin{equation}
\Gamma(z ; p, q) \equiv \prod_{i, j=0}^{\infty} \frac{1-p^{i+1} q^{j+1} z^{-1}}{1-p^i q^j z}
\end{equation}
where the series converges and by analytic continuation elsewhere. In this paper we will take $p=q=x$ in order to simplify the computation of the indices. Each of the indices presented in this paper can be further refined by considering different fugacities.
The check of a specific duality through the Lens space index consist in an identity between two such sum/integrals. We lack the tools to prove these identities analytically; what we do instead is to expand the indices in a Taylor series for small spacetime fugacities (small $x$) and match the indices order by order in this expansion. \\

The holonomies can be organized by uplifting them to the universal cover group $\tilde{G}$. The uplifted holonomies $\tilde{g}$ and $\tilde{h}$ satisfy:
\begin{equation} \label{eq:commutator_g_h_nu}
	\tilde{g} \cdot \tilde{h} \cdot \tilde{g}^{-1} \cdot \tilde{h}^{-1} = \nu
\end{equation}
\begin{equation} \label{eq:gr_mu}
	\tilde{g}^r = \mu
\end{equation}
where $\mu$ and $\nu$ range over the possible uplifts of $1 \in G$. For $\mathfrak{g}=\mathfrak{so}(2n+1)$ or $\mathfrak{usp}(2n)$ this means that there are four holonomy sectors $Z_{\mu,\nu}$ labelled by $(\mu,\nu)=(\pm 1, \pm 1)$. The Lens space index of the gauge theories with $\mathfrak{g}=\mathfrak{so}(2n+1)$ is \cite{Razamat:2013opa}:
\begin{equation}
\begin{aligned}
&\mathcal{I}_{S p i n\left(2n+1\right)}=Z_{1,1} \\
&\mathcal{I}_{S O\left(2n+1\right)_{+}}=\frac{1}{2}\left(Z_{1,1}+Z_{-1,1}+Z_{1,-1}+Z_{-1,-1}\right), \\
&\mathcal{I}_{S O\left(2n+1\right)_{-}}=\frac{1}{2}\left(Z_{1,1}+Z_{-1,1}+Z_{-1,-1}-Z_{1,-1}\right), \quad r=2 \bmod 4, \\
&\mathcal{I}_{S O\left(2n+1\right)_{-}}=\frac{1}{2}\left(Z_{1,1}+Z_{-1,1}+Z_{1,-1}-Z_{-1,-1}\right), \quad r=0 \bmod 4 .
\end{aligned}
\end{equation}

Similarly for $\mathfrak{g}=\mathfrak{usp}(2n)$:
\begin{equation}
\begin{aligned}
&\mathcal{I}_{USp\left(2n\right)}=Z_{1,1} \\
&\mathcal{I}_{\left(USp\left(2n\right)/\mathbb{Z}_2\right)_0}=
	\frac{1}{2}\left(Z_{1,1}+Z_{-1,1}+Z_{1,-1}+Z_{-1,-1}\right), \\
&\mathcal{I}_{\left(USp\left(2n\right)/\mathbb{Z}_2\right)_1}=
	\frac{1}{2}\left(Z_{1,1}+Z_{-1,1}+Z_{-1,-1}-Z_{1,-1}\right), \quad r=2 \bmod 4, \\
&\mathcal{I}_{\left(USp\left(2n\right)/\mathbb{Z}_2\right)_1}=
	\frac{1}{2}\left(Z_{1,1}+Z_{-1,1}+Z_{1,-1}-Z_{-1,-1}\right), \quad r=0 \bmod 4 .
\end{aligned}
\end{equation}


\section{$\mathcal{N}=4$ S-duality with $USp(2\nn )$ gauge group}
\label{sec:N4}

S-duality maps $\mathcal{N}=4$ SYM with gauge algebra $\mathfrak{g}$ and gauge coupling $\tau$ to SYM with gauge algebra $\mathfrak{g}'$ and gauge coupling $\tau' $. The dual algebra $\mathfrak{g}'$ can be either $\mathfrak{g}$ itself or the GNO dual algebra $\mathfrak{g}^\vee$. In this paper we consider the S-duality orbits for $\mathcal{N}=4$ SYM with gauge algebras $B_n= \mathfrak{so}(2n+1)$ and $C_n=\mathfrak{usp}(2n)$\footnote{We use notations where $USp(2) \simeq SU(2)$.}. The full duality group is the subgroup of $SL(2,\mathbb{R})$ generated by:
\begin{equation}
T^{\prime}=\left(\begin{array}{ll}
1 & 1 \\
0 & 1
\end{array}\right) \quad \text { and } \quad S^{\prime}=\left(\begin{array}{cc}
0 & 1 / \sqrt{2} \\
-\sqrt{2} & 0
\end{array}\right)
\end{equation}
which act on the gauge coupling $\tau$ by modular linear transformation:
\begin{equation}
	\left( \begin{array}{ll}
		a & b \\
		c & d
		\end{array}\right)
		:
		\tau \to \tau' = \frac{\tau a + b}{\tau c + d}
\end{equation}
Additionally, the $S'$ element of the duality group maps a theory with $\mathfrak{g}=\mathfrak{so}(2n+1)$ to a theory with $\mathfrak{g}' = \mathfrak{g}^\vee = \mathfrak{usp}(2n)$ while the $T'$ element of the duality group leaves the gauge algebra invariant.\\

\begin{figure}
\centering
	\includegraphics[scale=0.65]{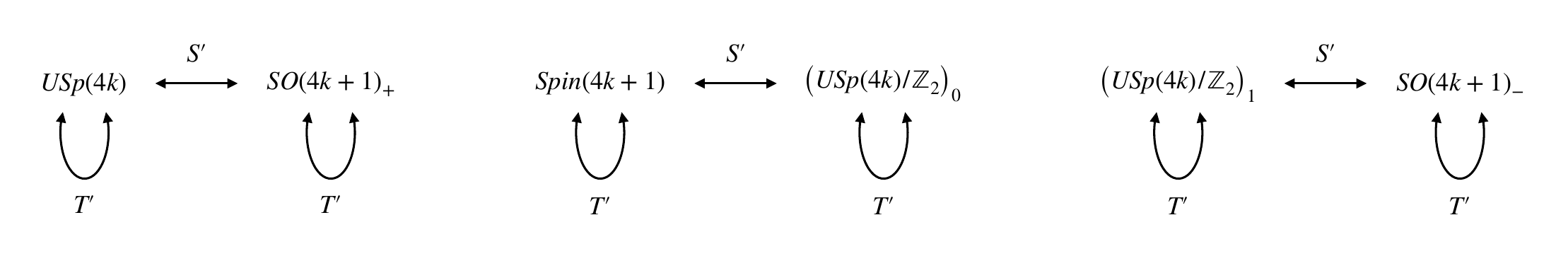}
	\caption{The S-duality orbits of $\mathcal{N}=4$ SYM with gauge algebra $\mathfrak{usp}(4k)$ and $\mathfrak{so}(4k+1)$, reproduced from \cite{Bergman:2022otk, Aharony:2013hda}.}
	\label{fig:even_n_orbits}
\end{figure}

\begin{figure}
\centering
	\includegraphics[scale=0.65]{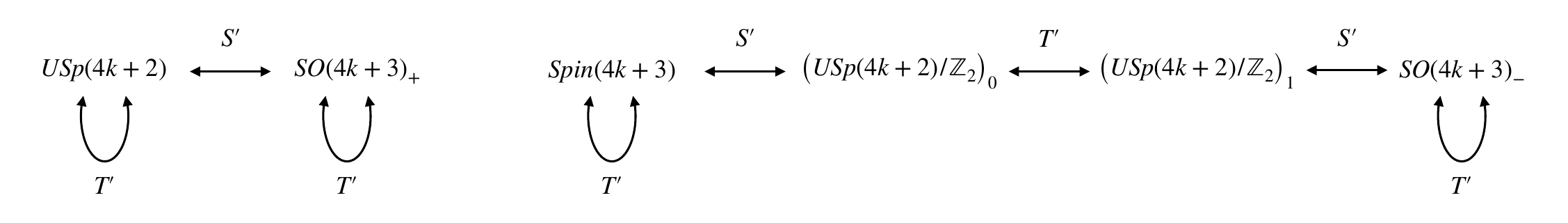}
	\caption{The S-duality orbits of $\mathcal{N}=4$ SYM with gauge algebra $\mathfrak{usp}(4k+2)$ and $\mathfrak{so}(4k+3)$, reproduced from \cite{Bergman:2022otk, Aharony:2013hda}.}
	\label{fig:odd_n_orbits}
\end{figure}

There are three choices for the global structure of both theories parametrized by the choice of gauge group and a discrete theta-like parameter. Borrowing the notation of \cite{Razamat:2013opa, Aharony:2013hda} they are:
\begingroup
\renewcommand{\arraystretch}{1.5}
\begin{equation}
\begin{array}{cccc}
\mathfrak{so}(2n+1): \quad&\quad Spin(2n+1)  \quad&\quad  SO(2n+1)_+   \quad&\quad SO(2n+1)_-
\\
\mathfrak{usp}(2n):  \quad&\quad  USp(2n)  \quad&\quad  (USp(2n)/ \mathbb{Z}_2)_0  \quad&\quad   (USp(2n)/ \mathbb{Z}_2)_1
\end{array}
\end{equation}
\endgroup
where $SO(2n+1)_+ = (Spin(2n+1)/ \mathbb{Z}_2)_0$ and $SO(2n+1)_- =  (Spin(2n+1)/ \mathbb{Z}_2)_1$. The S-duality group forms different orbits depending on whether $n$ is even (Figure \ref{fig:even_n_orbits}) or odd (Figure \ref{fig:odd_n_orbits}). In this section we perform a precision test of the S-duality orbits by computing the Lens space index of these theories for low values of $n$ and small fugacities. We will see that the index is the same between theories that lie in the same orbit and is different between theories in different orbits.

\subsection{$SO(5)$ and $USp(4)$}
At the level of the gauge algebra we have $\mathfrak{so}(5) \sim \mathfrak{usp}(4)$, while at the level of the gauge groups we have the isomorphisms:
\begin{equation}
\begin{split}
	Spin(5) \sim& \,USp(4) \\
	SO(5) \sim& \,USp(4)/\mathbb{Z}_2
\end{split}.
\end{equation}

\begin{figure}	
\centering
\includegraphics[scale=0.8]{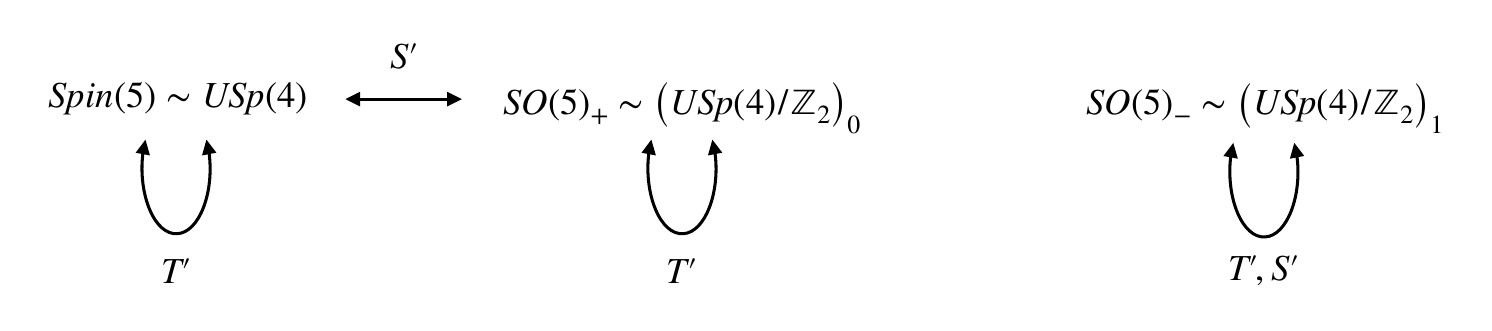}
\caption{S-duality orbits of $\mathcal{N}=4$ SYM with gauge algebra $\mathfrak{usp}(4)=\mathfrak{so}(4)$.}
\label{eq:spin5orbits}
\end{figure}

The S-duality orbits reduce to the ones in Figure \ref{eq:spin5orbits}.
We see that we still have a non-trivial prediction from S-duality which implies:
\begin{equation}
	\mathcal{I}_{Spin(5)} = \mathcal{I}_{SO(5)_+}
\end{equation}

\subsubsection*{r=2}
The sectors $Z_{\mu,\nu}$ contributes to the $\mathfrak{so}(5) \sim \mathfrak{usp}(4)$ theory up to $\mathcal{O}(x^6)$ as:

\munutable{\mathcal{N}=4 \quad \mathfrak{so}(5) \sim \mathfrak{usp}(4)}{2}
{3+24 x^{4/3}-31 x^2+120 x^{8/3}-\\-210 x^{10/3}+326 x^4-312 x^{14/3}+\\+219 x^{16/3}+1330 x^6}
{1+3 x^{2/3}+9 x^{4/3}+13 x^2+6 x^{8/3}+\\+18 x^{10/3}-58 x^4+159 x^{14/3}+\\+111 x^{16/3}-122 x^6}
{1+x^2-6 x^{8/3}+36 x^{10/3}-\\-80 x^4+96 x^{14/3}-27 x^{16/3}-256 x^6}
{1-3 x^{2/3}+15 x^{4/3}-45 x^2+\\+120 x^{8/3}-264 x^{10/3}+464 x^4-\\-567 x^{14/3}+135 x^{16/3}+1708 x^6}

The Lens space indices for the three global structures are:
\begin{equation}
\begin{split}
	\mathcal{I}_{Spin(5) } = \mathcal{I}_{SO(5)_+ } = \,&
		3+24 x^{4/3}-31 x^2+120 x^{8/3}-210 x^{10/3}+\\&+326 x^4-312 x^{14/3}+219 x^{16/3}+1330 x^6 
\end{split}
\end{equation}

\begin{equation}
\begin{split}
	\mathcal{I}_{SO(5)_-} = \,&2+24 x^{4/3}-32 x^2+126 x^{8/3}-246 x^{10/3}+\\&+406 x^4-408 x^{14/3}+246 x^{16/3}+1586 x^6
\end{split}
\end{equation}

\subsubsection*{r=4}
Up to $\mathcal{O}(x^2)$ the indices are:
\munutable{\mathcal{N}=4 \quad \mathfrak{so}(5) \sim \mathfrak{usp}(4)}{4}
{6+9 x^{2/3}+51 x^{4/3}+3 x^2}
{3+12 x^{2/3}+33 x^{4/3}+47 x^2}
{2-3 x^{2/3}+15 x^{4/3}-45 x^2}
{1+3 x^{4/3}+x^2}

\begin{equation}
	\mathcal{I}_{Spin(5)} = \mathcal{I}_{SO(5)_+ }  =\, 6+9 x^{2/3}+51 x^{4/3}+3 x^2
\end{equation}
\begin{equation}
	\mathcal{I}_{SO(5)_-} = 5+9 x^{2/3}+48 x^{4/3}+2 x^2
\end{equation}

\subsubsection*{r=6}
Up to $\mathcal{O}(x^2)$ the indices are:
\munutable{\mathcal{N}=4 \quad \mathfrak{so}(5) \sim \mathfrak{usp}(4)}{6}
{10+24 x^{2/3}+93 x^{4/3}+57 x^2}
{6+27 x^{2/3}+72 x^{4/3}+101 x^2}
{2+3 x^{4/3}+x^2}
{2-3 x^{2/3}+18 x^{4/3}-45 x^2}

\begin{equation}
	\mathcal{I}_{Spin(5)} = \mathcal{I}_{SO(5)_+ }  = 10+24 x^{2/3}+93 x^{4/3}+57 x^2
\end{equation}
\begin{equation}
	\mathcal{I}_{SO(5)_-} = 8+24 x^{2/3}+90 x^{4/3}+56 x^2
\end{equation}

\subsubsection*{r=8}
Up to $\mathcal{O}(x^2)$ the indices are:
\munutable{\mathcal{N}=4 \quad \mathfrak{so}(5) \sim \mathfrak{usp}(4)}{8}
{15+45 x^{2/3}+150 x^{4/3}+131 x^2}
{10+48 x^{2/3}+126 x^{4/3}+175 x^2}
{3-3 x^{2/3}+18 x^{4/3}-45 x^2}
{2+6 x^{4/3}+x^2}

\begin{equation}
	\mathcal{I}_{Spin(5)} = \mathcal{I}_{SO(5)_+ }  = 15+45 x^{2/3}+150 x^{4/3}+131 x^2
\end{equation}
\begin{equation}
	\mathcal{I}_{SO(5)_-} = 13+45 x^{2/3}+144 x^{4/3}+130 x^2
\end{equation}

\subsection{$SO(7)$ and $USp(6)$}
This is the lowest rank where S-duality relates theories with different gauge algebras. The S-duality orbits are given in Figure \ref{fig:odd_n_orbits} with $k=1$. 

\subsubsection*{r=2}
The contributions from the $Z_{\mu,\nu}$ sectors for the theories with gauge algebras $\mathfrak{so}(7)$ and $\mathfrak{usp}(6)$ up to $\mathcal{O}(x^4)$ are:
\munutable{\mathcal{N}=4 \quad \mathfrak{so}(7)}{2}
{3+30 x^{4/3}-39 x^2+210 x^{8/3}-\\-411 x^{10/3}+1146 x^4}
{2+3 x^{2/3}+15 x^{4/3}+16 x^2+\\+39 x^{8/3}+69 x^{10/3}-56 x^4}
{1+6 x^{4/3}-7 x^2+12 x^{8/3}+\\+9 x^{10/3}-60 x^4}
{2-3 x^{2/3}+21 x^{4/3}-62 x^2+\\+183 x^{8/3}-471 x^{10/3}+1142 x^4}
\munutable{\mathcal{N}=4 \quad \mathfrak{usp}(6)}{2}
{4+36 x^{4/3}-46 x^2+222 x^{8/3}-\\-402 x^{10/3}+1086 x^4}
{1+3 x^{2/3}+9 x^{4/3}+23 x^2+\\+27 x^{8/3}+60 x^{10/3}+4 x^4}
{0}
{1-3 x^{2/3}+15 x^{4/3}-55 x^2+\\+171 x^{8/3}-480 x^{10/3}+1202 x^4}

The Lens space indices for the six possible theories are:
\begin{equation}
\begin{split}
	\mathcal{I}_{Spin(7)} =&\mathcal{I}_{SO(7)_-} =\mathcal{I}_{(USp(6)/\mathbb{Z}_2)_0} =\mathcal{I}_{(USp(6)/\mathbb{Z}_2)_1} = \\
	&= 3+30 x^{4/3}-39 x^2+210 x^{8/3}-411 x^{10/3}+1146 x^4
\end{split}
\end{equation}
\begin{equation}
	\mathcal{I}_{SO(7)_+} = \mathcal{I}_{USp(6)} =4+36 x^{4/3}-46 x^2+222 x^{8/3}-402 x^{10/3}+1086 x^4
\end{equation}
We see that the Lens space index of theories that lie in the same orbit matches up to this order while the indices of theories that lie in different orbits are different. 

\subsubsection*{r=4}
Up to $\mathcal{O}(x^2)$ the indices are:
\munutable{\mathcal{N}=4 \quad \mathfrak{so}(7)}{4}
{8+15 x^{2/3}+96 x^{4/3}+68 x^2}
{6+21 x^{2/3}+72 x^{4/3}+142 x^2}
{4-3 x^{2/3}+36 x^{4/3}-78 x^2}
{2+3 x^{2/3}+12 x^{4/3}-4 x^2}
\munutable{\mathcal{N}=4 \quad \mathfrak{usp}(6)}{4}
{10+18 x^{2/3}+108 x^{4/3}+64 x^2}
{4+18 x^{2/3}+60 x^{4/3}+146 x^2}
{2-6 x^{2/3}+24 x^{4/3}-74 x^2}
{0}

\begin{equation}
	\mathcal{I}_{Spin(7)} =\mathcal{I}_{SO(7)_-} =\mathcal{I}_{(USp(6)/\mathbb{Z}_2)_0} =\mathcal{I}_{(USp(6)/\mathbb{Z}_2)_1} =  8+15 x^{2/3}+96 x^{4/3}+68 x^2
\end{equation}
\begin{equation}
	\mathcal{I}_{SO(7)_+} = \mathcal{I}_{USp(6)} =10+18 x^{2/3}+108 x^{4/3}+64 x^2
\end{equation}

\subsubsection*{r=6}
Up to $\mathcal{O}(x^2)$ the indices are:
\munutable{\mathcal{N}=4 \quad \mathfrak{so}(7)}{6}
{16+54 x^{2/3}+240 x^{4/3}+369 x^2}
{14+60 x^{2/3}+213 x^{4/3}+452 x^2}
{4+6 x^{2/3}+24 x^{4/3}-x^2}
{6+51 x^{4/3}-84 x^2}
\munutable{\mathcal{N}=4 \quad \mathfrak{usp}(6)}{6}
{20+60 x^{2/3}+264 x^{4/3}+368 x^2}
{10+54 x^{2/3}+189 x^{4/3}+453 x^2}
{0}
{2-6 x^{2/3}+27 x^{4/3}-83 x^2}

\begin{equation}
	\mathcal{I}_{Spin(7)} =\mathcal{I}_{SO(7)_-} =\mathcal{I}_{(USp(6)/\mathbb{Z}_2)_0} =\mathcal{I}_{(USp(6)/\mathbb{Z}_2)_1} =  16+54 x^{2/3}+240 x^{4/3}+369 x^2
\end{equation}
\begin{equation}
	\mathcal{I}_{SO(7)_+} = \mathcal{I}_{USp(6)} =20+60 x^{2/3}+264 x^{4/3}+368 x^2
\end{equation}

\subsection{$SO(9)$ and $USp(8)$}
The duality orbits are given in Figure \ref{fig:even_n_orbits} with $k=2$.

\subsubsection*{r=2}
Up to $\mathcal{O}(x^2)$ the indices are:
\munutable{\mathcal{N}=4 \quad \mathfrak{so}(9)}{2}
{4+36 x^{4/3}-46 x^2}
{2+3 x^{2/3}+21 x^{4/3}+8 x^2}
{2+12 x^{4/3}-14 x^2}
{2-3 x^{2/3}+27 x^{4/3}-70 x^2}
\munutable{\mathcal{N}=4 \quad \mathfrak{usp}(8)}{2}
{5+48 x^{4/3}-61 x^2}
{1+3 x^{2/3}+9 x^{4/3}+23 x^2}
{1+x^2}
{1-3 x^{2/3}+15 x^{4/3}-55 x^2}

\begin{equation}
	\mathcal{I}_{Spin(9)} =\mathcal{I}_{(USp(8)/\mathbb{Z}_2)_0} =  4+36 x^{4/3}-46 x^2
\end{equation}
\begin{equation}
	\mathcal{I}_{SO(9)_+} =\mathcal{I}_{USp(8)} =  5+48 x^{4/3}-61 x^2
\end{equation}
\begin{equation}
	\mathcal{I}_{SO(9)_-} =\mathcal{I}_{(USp(8)/\mathbb{Z}_2)_1} =  3+36 x^{4/3}-47 x^2
\end{equation}

\subsubsection*{r=4}
Up to $\mathcal{O}(x^2)$ the indices are:
\munutable{\mathcal{N}=4 \quad \mathfrak{so}(9)}{4}
{12+24 x^{2/3}+153 x^{4/3}+157 x^2}
{8+30 x^{2/3}+120 x^{4/3}+256 x^2}
{6+63 x^{4/3}-89 x^2}
{4+6 x^{2/3}+36 x^{4/3}+12 x^2}
\munutable{\mathcal{N}=4 \quad \mathfrak{usp}(8)}{4}
{15+30 x^{2/3}+186 x^{4/3}+168 x^2}
{5+24 x^{2/3}+87 x^{4/3}+245 x^2}
{3-6 x^{2/3}+30 x^{4/3}-100 x^2}
{1+3 x^{4/3}+x^2}

\begin{equation}
	\mathcal{I}_{Spin(9)} =\mathcal{I}_{(USp(8)/\mathbb{Z}_2)_0} =  12+24 x^{2/3}+153 x^{4/3}+157 x^2
\end{equation}
\begin{equation}
	\mathcal{I}_{SO(9)_+} =\mathcal{I}_{USp(8)} =  15+30 x^{2/3}+186 x^{4/3}+168 x^2
\end{equation}
\begin{equation}
	\mathcal{I}_{SO(9)_-} =\mathcal{I}_{(USp(8)/\mathbb{Z}_2)_1} =  11+24 x^{2/3}+150 x^{4/3}+156 x^2
\end{equation}

\subsubsection*{r=6}
Up to $\mathcal{O}(x^2)$ the indices are:
\munutable{\mathcal{N}=4 \quad \mathfrak{so}(9)}{6}
{28+102 x^{2/3}+486 x^{4/3}+1014 x^2}
{22+108 x^{2/3}+444 x^{4/3}+1121 x^2}
{10+18 x^{2/3}+90 x^{4/3}+62 x^2}
{10+12 x^{2/3}+120 x^{4/3}-49 x^2}
\munutable{\mathcal{N}=4 \quad \mathfrak{usp}(8)}{6}
{35+120 x^{2/3}+570 x^{4/3}+1074 x^2}
{15+90 x^{2/3}+360 x^{4/3}+1061 x^2}
{3+6 x^{4/3}+2 x^2}
{3-6 x^{2/3}+36 x^{4/3}-109 x^2}

\begin{equation}
	\mathcal{I}_{Spin(9)} =\mathcal{I}_{(USp(8)/\mathbb{Z}_2)_0} =  28+102 x^{2/3}+486 x^{4/3}+1014 x^2
\end{equation}
\begin{equation}
	\mathcal{I}_{SO(9)_+} =\mathcal{I}_{USp(8)} =  35+120 x^{2/3}+570 x^{4/3}+1074 x^2
\end{equation}
\begin{equation}
	\mathcal{I}_{SO(9)_-} =\mathcal{I}_{(USp(8)/\mathbb{Z}_2)_1} =  25+102 x^{2/3}+480 x^{4/3}+1012 x^2
\end{equation}

\subsection{$SO(11)$ and $USp(10)$}
The duality orbits are given in Figure \ref{fig:odd_n_orbits} with $k=2$.

\subsubsection*{r=2}
Up to $\mathcal{O}(x)$ the indices are:
\munutable{\mathcal{N}=4 \quad \mathfrak{so}(11)}{2}
{4 \qquad \,}
{3+3x^{2/3}}
{2}
{3-3x^{2/3}}
\munutable{\mathcal{N}=4 \quad \mathfrak{usp}(10)}{6}
{6 \qquad \,}
{1+x^{2/3}}
{0}
{1-x^{2/3}}

\begin{equation}
	\mathcal{I}_{Spin(11)} =\mathcal{I}_{SO(11)_-} =\mathcal{I}_{(USp(10)/\mathbb{Z}_2)_0} =\mathcal{I}_{(USp(10)/\mathbb{Z}_2)_1} =  4
\end{equation}
\begin{equation}
	\mathcal{I}_{SO(11)_+} = \mathcal{I}_{USp(10)} =6
\end{equation}

%
%
%
\section{$\mathcal{N}=2$ S-duality of elliptic models with orientifolds}
\label{sec:N2}
In this section we consider $\mathcal{N}=2$ quiver theories that contain orthogonal or symplectic gauge groups. This theories were studied in \cite{Uranga:1998uj} where they arose as low energy gauge theories of elliptic models with orientifolds in Type IIA string theory. The brane construction involves $D4$ branes wrapped on a compact dimension as well as $NS5$ branes and $O6^\pm$ orientifold planes. The configuration preserves $d=4$ $\mathcal{N}=2$ supersymmetry and the low energy theory on the stack of $D4$ branes is a quiver gauge theory. Due to the presence of the orientifold planes the quiver can contain real (orthogonal and/or symplectic) gauge groups as well as two-index tensorial representations of unitary gauge groups. These theories have a $\mathbb{Z}_2$ subgroup of the center that is not broken by the matter content, therefore we can consider different global structures. In particular there are three choices of global structures, analogously to the case of $\mathcal{N}=4$ with gauge algebras $\mathfrak{so}(2n+1)$ and $\mathfrak{usp}(2n)$ studied in the previous section.\\

In this paper we consider the case of quivers with two gauge nodes of the families $ii)$ and $iii)$ of  \cite{Uranga:1998uj}. The quivers in $\mathcal{N}=1$ notation are shown in \eqref{eq:N2quivers}.
\begin{equation}	\label{eq:N2quivers}
\includegraphics[scale=0.6]{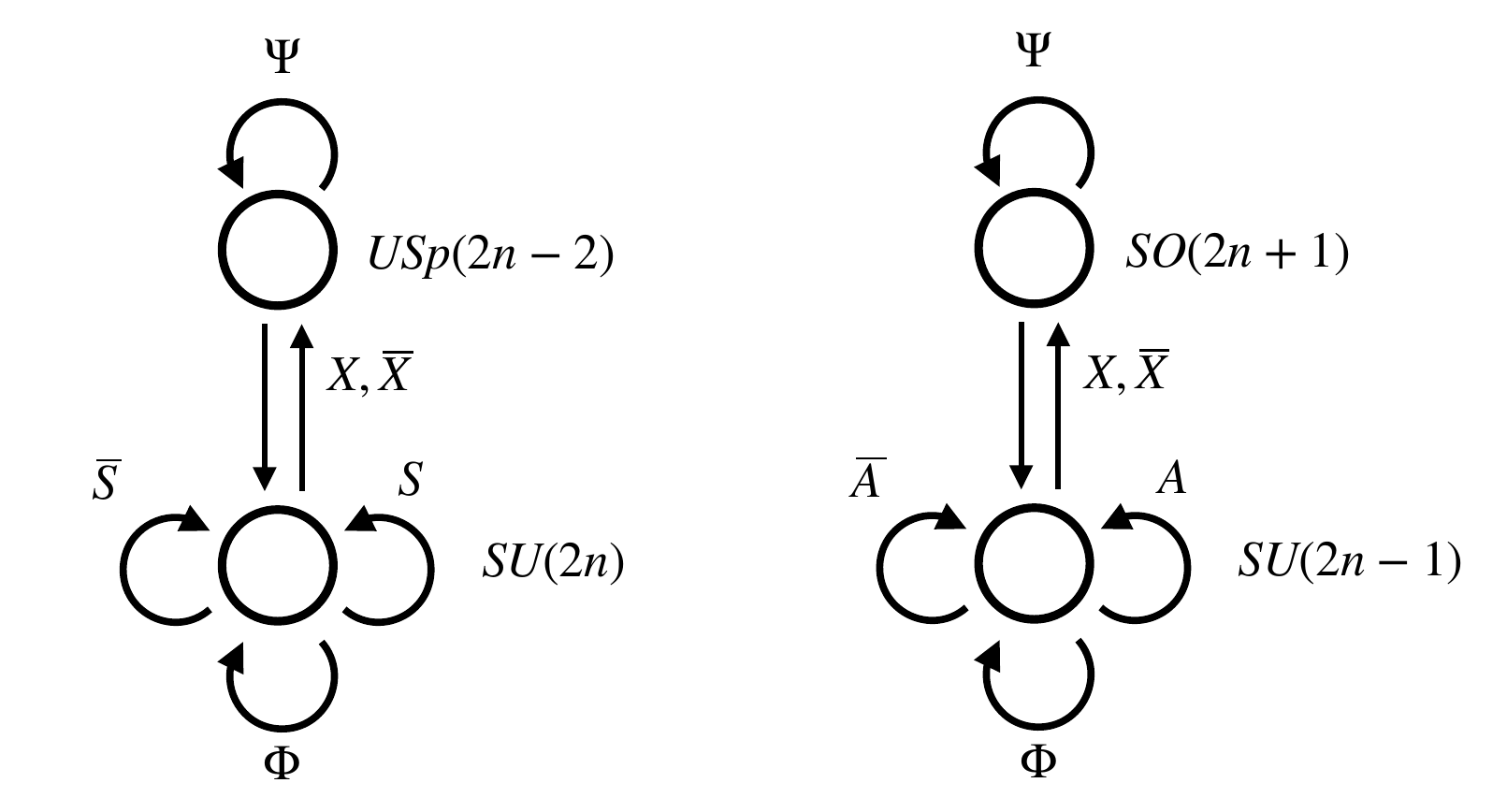}
\end{equation}

 The fields have the following representations:
\begin{equation}
\renewcommand{\arraystretch}{1.3}
\begin{array}{|c|c|c|c|}
\hline \mathcal{T}^{(SO)} & \mathfrak{so}(2n+1) & \mathfrak{su}(2n-1) & R \\
\hline \Psi & adj & 1 & 2 \\
\hline \Phi & 1 & adj & 2 \\
\hline A & 1 & {\tiny\Yvcentermath1\yng(1,1)} & \frac{2}{3} \\
\hline \overline{A} & 1 & {\tiny\Yvcentermath1\overline{\yng(1,1)}} & \frac{2}{3} \\
\hline X & vec &  {\tiny\Yvcentermath1\yng(1)} & \frac{2}{3} \\
\hline \overline{X} & vec &  {\tiny\Yvcentermath1\overline{\yng(1)}} & \frac{2}{3} \\
\hline
\end{array}
\qquad
\begin{array}{|c|c|c|c|}
\hline \mathcal{T}^{(USp)} & \mathfrak{usp}(2n-2) & \mathfrak{su}(2n) & R \\
\hline \Psi & adj & 1 & 2 \\
\hline \Phi & 1 & adj & 2 \\
\hline S & 1 & {\tiny\Yvcentermath1\yng(2)} & \frac{2}{3} \\
\hline \overline{S} & 1 & {\tiny\Yvcentermath1\overline{\yng(2)}} & \frac{2}{3} \\
\hline X & fund &  {\tiny\Yvcentermath1\yng(1)} & \frac{2}{3} \\
\hline \overline{X} & fund &  {\tiny\Yvcentermath1\overline{\yng(1)}} & \frac{2}{3} \\
\hline
\end{array}
\renewcommand{\arraystretch}{1}
\end{equation}
There are three choices of global form for each of these theories, namely:
\begin{equation}
\label{eq:N2_SO_thys}
\begin{split}
 \mathcal{T}^{(Spin)} (n)=& Spin(2n+1) \times SU(2n-1)\\
 \mathcal{T}^{(SO)}_+(n) =& (Spin(2n+1)/\mathbb{Z}_2)_0 \times SU(2n-1)\\
 \mathcal{T}^{(SO)}_- (n)=&(Spin(2n+1)/\mathbb{Z}_2)_1 \times SU(2n-1)\\
\end{split}
\end{equation}
and:
\begin{equation}
\label{eq:N2_USp_thys}
\begin{split}
\mathcal{T}^{(USp)}(n) =& USp(2n-2) \times SU(2n)\\
\mathcal{T}^{(USp)}_+(n) =&\left( \left(USp(2n-2) \times SU(2n) \right)/\mathbb{Z}_2\right)_0\\
\mathcal{T}^{(USp)}_- (n)=&\left( \left(USp(2n-2) \times SU(2n) \right)/\mathbb{Z}_2\right)_1\\
\end{split}	
\end{equation}

\begin{figure}	
	\centering
	\includegraphics[scale=0.7]{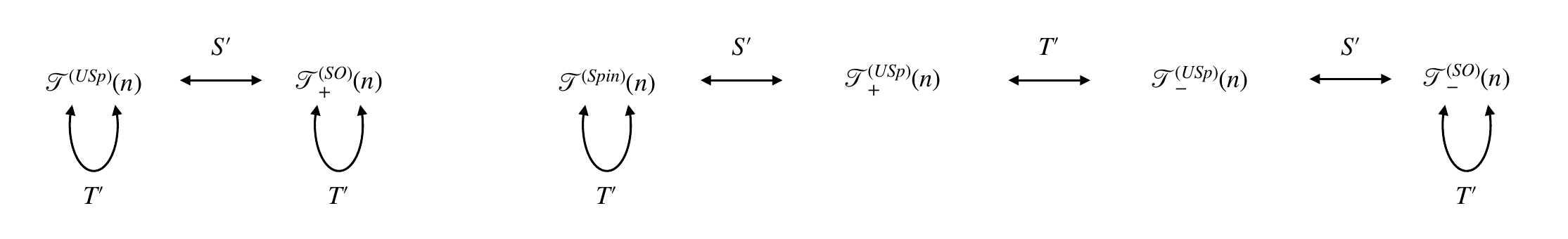}
	\caption{S-duality orbits of the $\mathcal{N}=2$ and $\mathcal{N}=1$ gauge theories in \eqref{eq:N2quivers} and \eqref{eq:N1quivers} respectively.}
	\label{eq:N2orbits}
\end{figure}

In principle these theories can be engineered in Type IIB by probing the non-chiral $\mathbb{C}^3/\mathbb{Z}_3$ orbifold with $2n$ $D3$ branes and introducing an $O3$ plane on top of the stack of $D3$ branes.  The number of $D3$ branes plus their orbifold images is $6n$, therefore we expect that the S-duality orbits will be the same as the ones of $\mathcal{N}=4$ SYM with gauge algebras $\mathfrak{usp}(6n)$ and $\mathfrak{so}(6n+1)$ shown in Figure \ref{eq:N2orbits}.\\

The holonomy sectors can be organized by $(\mu,\nu)=(\pm1, \pm1)$. We call the contributions from such sectors $Z^{(USp)}_{\mu,\nu}$ and $Z^{(Spin)}_{\mu,\nu}$ for $\mathcal{T}^{(USp)}$ and $\mathcal{T}^{(Spin)}$ respectively. The  indices of the two theories are given by:

\begin{equation}
\begin{split}
	\mathcal{I}_{\mathcal{T}^{(USp)}(n)} =& \,  Z^{(USp)}_{1,1} \\
	\mathcal{I}_{\mathcal{T}^{(USp)}_+ (n)} =& \,  Z^{(USp)}_{1,1}+ Z^{(USp)}_{-1,1}+ Z^{(USp)}_{1,-1}+ Z^{(USp)}_{-1,-1} \\
	\mathcal{I}_{\mathcal{T}^{(USp)}_- (n)} =& \,  Z^{(USp)}_{1,1}+ Z^{(USp)}_{-1,1} + Z^{(USp)}_{-1,-1} - Z^{(USp)}_{1,-1}, \qquad r=2 \text{ mod } 4,\\
	\mathcal{I}_{\mathcal{T}^{(USp)}_- (n)} =& \,  Z^{(USp)}_{1,1}+ Z^{(USp)}_{-1,1} + Z^{(USp)}_{1,-1} - Z^{(USp)}_{-1,-1}, \qquad r=0 \text{ mod } 4,\\
\end{split}
\end{equation}

\begin{equation}
\begin{split}
	\mathcal{I}_{\mathcal{T}^{(Spin)}(n)} =& \,  Z^{(Spin)}_{1,1} \\
	\mathcal{I}_{\mathcal{T}^{(SO)}_+ (n)} =& \,  Z^{(Spin)}_{1,1}+ Z^{(Spin)}_{-1,1}+ Z^{(Spin)}_{1,-1}+ Z^{(Spin)}_{-1,-1} \\
	\mathcal{I}_{\mathcal{T}^{(SO)}_- (n)} =& \,  Z^{(Spin)}_{1,1}+ Z^{(Spin)}_{-1,1} + Z^{(Spin)}_{-1,-1} - Z^{(Spin)}_{1,-1}, \qquad r=2 \text{ mod } 4,\\
	\mathcal{I}_{\mathcal{T}^{(SO)}_- (n)} =& \,  Z^{(Spin)}_{1,1}+ Z^{(Spin)}_{-1,1} + Z^{(Spin)}_{1,-1} - Z^{(Spin)}_{-1,-1}, \qquad r=0 \text{ mod } 4,\\
\end{split}
\end{equation}

In this section we compute the Lens space index for these theories with $n=2$ and $r=2,4$ and with $n=3$ and $r=2$. We find that the indices match for theories that lie in the same orbit and is different between theories that lie in different orbits.

\subsection{$SO(5)\times SU(3)$ and $USp(2)\times SU(4)$}
The smallest value of $n$ for which all the groups have positive ranks is $n=2$. Then the gauge algebra for the two theories are $\mathfrak{so}(5) \times \mathfrak{su}(3)$ and $\mathfrak{usp}(2) \times \mathfrak{su}(4)$.  We notice that for this value of $n$ we could regard $\mathfrak{usp}(2)$ as $\mathfrak{su}(2)$ and all the gauge groups would be either orthogonal or special unitary. The index could be computed without the technology developed in this paper for symplectic gauge groups. This is only true for $n=2$, while for higher $n$ the technology for computing the index in the presence of symplectic gauge groups is needed.

\subsubsection*{r=2}
The contributions to the indices up to $\mathcal{O}(x^2)$ are:
\munutable{\mathcal{N}=2 \quad \mathfrak{so}(5)\times \mathfrak{su}(3)}{2}
{3+x^{2/3}+19 x^{4/3}-x^2}
{2+3 x^{2/3}+11 x^{4/3}+21 x^2}
{1+x^{2/3}+3 x^{4/3}+3 x^2}
{2-x^{2/3}+11 x^{4/3}-19 x^2}

\munutable{\mathcal{N}=2 \quad \mathfrak{usp}(2)\times \mathfrak{su}(4)}{2}
{4+2 x^{2/3}+22 x^{4/3}+2 x^2}
{1+2 x^{2/3}+8 x^{4/3}+18 x^2}
{0}
{1-2 x^{2/3}+8 x^{4/3}-22 x^2}

The indices are:
\begin{equation}
	\mathcal{I}_{\mathcal{T}^{(Spin)}(2)} = \mathcal{I}_{\mathcal{T}^{(USp)}_+ (2)} = \mathcal{I}_{\mathcal{T}^{(USp)}_- (2)} = \mathcal{I}_{\mathcal{T}^{(SO)}_- (2)} = 3+x^{2/3}+19 x^{4/3}-x^2
\end{equation}
\begin{equation}
	\mathcal{I}_{\mathcal{T}^{(USp)}(2)} = \mathcal{I}_{\mathcal{T}^{(SO)}_+ (2)} = 4+2 x^{2/3}+22 x^{4/3}+2 x^2
\end{equation}

\subsubsection*{r=4}
The contributions to the indices up to $\mathcal{O}(x^2)$ are:
\munutable{\mathcal{N}=2 \quad \mathfrak{so}(5)\times \mathfrak{su}(3)}{4}
{8+13 x^{2/3}+68 x^{4/3}+81 x^2}
{6+17 x^{2/3}+55 x^{4/3}+111 x^2}
{4+x^{2/3}+22 x^{4/3}-17 x^2}
{2+5 x^{2/3}+9 x^{4/3}+13 x^2}

\munutable{\mathcal{N}=2 \quad \mathfrak{usp}(2)\times \mathfrak{su}(4)}{4}
{10+18 x^{2/3}+77 x^{4/3}+94 x^2}
{4+12 x^{2/3}+46 x^{4/3}+98 x^2}
{2-4 x^{2/3}+13 x^{4/3}-30 x^2}
{0}

The indices are:
\begin{equation}
	\mathcal{I}_{\mathcal{T}^{(Spin)}(2)} = \mathcal{I}_{\mathcal{T}^{(USp)}_+ (2)} = \mathcal{I}_{\mathcal{T}^{(USp)}_- (2)} = \mathcal{I}_{\mathcal{T}^{(SO)}_- (2)} = 8+13 x^{2/3}+68 x^{4/3}+81 x^2 \end{equation}
\begin{equation}
	\mathcal{I}_{\mathcal{T}^{(USp)}(2)} = \mathcal{I}_{\mathcal{T}^{(SO)}_+ (2)} = 
	10+18 x^{2/3}+77 x^{4/3}+94 x^2
\end{equation}

\subsection{$SO(7)\times SU(5)$ and $USp(4)\times SU(6)$}
In this section we consider the case of $n=3$. Then the gauge algebras are $\mathfrak{so}(7)\times \mathfrak{su}(5)$ and $\mathfrak{usp}(4)\times \mathfrak{su}(6)$.

\subsubsection*{r=2}
The contributions to the indices up to $\mathcal{O}(x^2)$ are:
\munutable{\mathcal{N}=2 \quad \mathfrak{so}(7)\times \mathfrak{su}(5)}{2}
{4+2 x^{2/3}+27 x^{4/3}+10 x^2}
{3+4 x^{2/3}+20 x^{4/3}+29 x^2}
{2+2 x^{2/3}+13 x^{4/3}+8 x^2}
{3+20 x^{4/3}-11 x^2}

\munutable{\mathcal{N}=2 \quad \mathfrak{usp}(4)\times \mathfrak{su}(6)}{2}
{6+4 x^{2/3}+40 x^{4/3}+18 x^2}
{1+2 x^{2/3}+7 x^{4/3}+21 x^2}
{0}
{1-2 x^{2/3}+7 x^{4/3}-19 x^2}

The indices are:
\begin{equation}
	\mathcal{I}_{\mathcal{T}^{(Spin)}(4)} = \mathcal{I}_{\mathcal{T}^{(USp)}_+ (4)} = \mathcal{I}_{\mathcal{T}^{(USp)}_- (4)} = \mathcal{I}_{\mathcal{T}^{(SO)}_- (4)} = 4 + 2 x^{2/3} + 27 x^{4/3} + 10 x^2
\end{equation}
\begin{equation}
	\mathcal{I}_{\mathcal{T}^{(USp)}(4)} = \mathcal{I}_{\mathcal{T}^{(SO)}_+ (4)} =6+4 x^{2/3}+40 x^{4/3}+18 x^2
\end{equation}

%
%
%
\section{$\mathcal{N}=1$ inherited S-duality of elliptic models with orientifolds}
\label{sec:N1}
In this section we consider $\mathcal{N}=1$ models that can be obtained from the models studied in Section \ref{sec:N2} by adding a mass deformation for the adjoints. The mass deformation breaks supersymmetry to $\mathcal{N}=1$. When the massive adjoints are integrated out we obtain the gauge theories described by the quivers in \eqref{eq:N1quivers}.
\begin{equation}	\label{eq:N1quivers}
\includegraphics[scale=0.6]{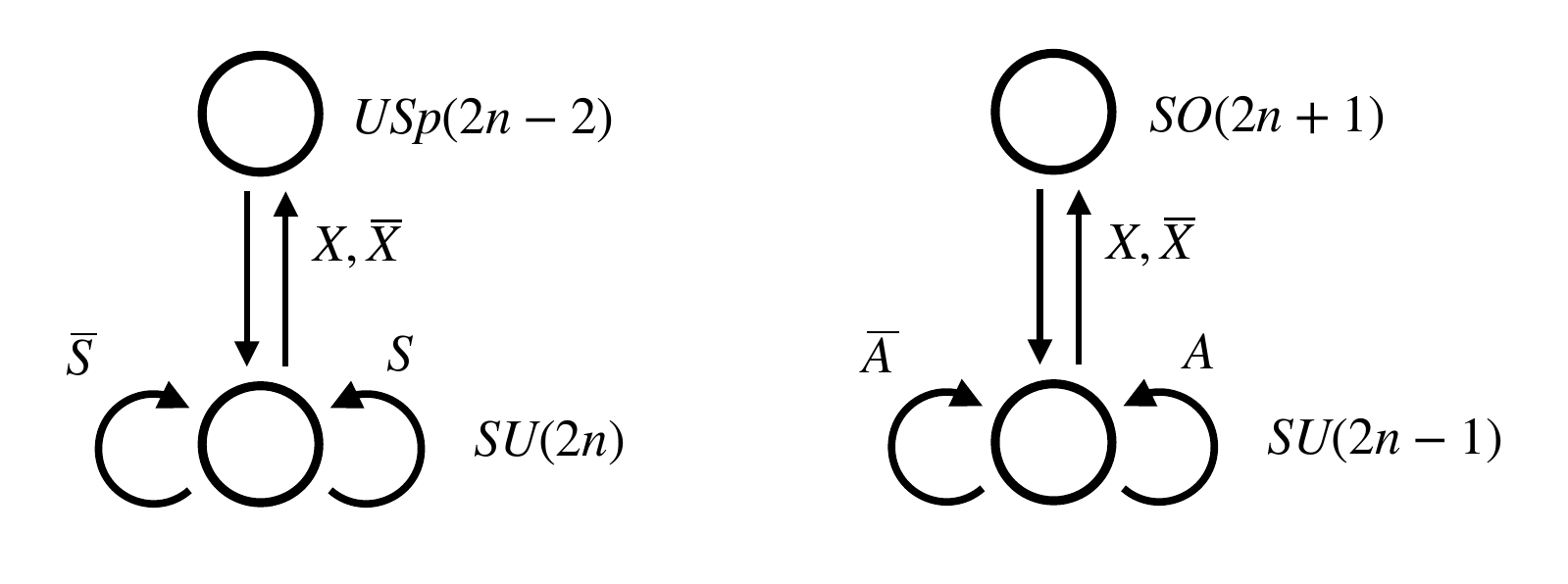}
\end{equation}

The superpotentials are, respectively:
\begin{equation}
\begin{split}
	W^{(USp)} =&\, X \overline{X} S \overline{S} \\
	W^{(SO)} =&\, X \overline{X} A \overline{A} 
\end{split}
\end{equation}

The fields have the following representations:
\begin{equation}
\renewcommand{\arraystretch}{1.3}
\begin{array}{|c|c|c|c|}
\hline \mathcal{T}^{(SO)} & \mathfrak{so}(2n+1) & \mathfrak{su}(2n-1) &  R  \\
\hline A & 1 & {\tiny\Yvcentermath1\yng(1,1)} & \frac{1}{2} \\
\hline \overline{A} & 1 & {\tiny\Yvcentermath1\overline{\yng(1,1)}} & \frac{1}{2} \\
\hline X & vec &  {\tiny\Yvcentermath1\yng(1)} & \frac{1}{2} \\
\hline \overline{X} & vec &  {\tiny\Yvcentermath1\overline{\yng(1)}} &  \frac{1}{2} \\
\hline
\end{array}
\qquad
\begin{array}{|c|c|c|c|}
\hline \mathcal{T}^{(USp)} & \mathfrak{usp}(2n-2) & \mathfrak{su}(2n) & R \\
\hline S & 1 & {\tiny\Yvcentermath1\yng(2)} & \frac{1}{2} \\
\hline \overline{S} & 1 & {\tiny\Yvcentermath1\overline{\yng(2)}} &  \frac{1}{2} \\
\hline X & fund &  {\tiny\Yvcentermath1\yng(1)} &  \frac{1}{2} \\
\hline \overline{X} & fund &  {\tiny\Yvcentermath1\overline{\yng(1)}} &  \frac{1}{2} \\
\hline
\end{array}
\renewcommand{\arraystretch}{1}
\end{equation}

A large family of theories that includes these models where first studied in \cite{Antinucci:2021edv}, arising from orbifold projection of toric theories. The authors also presented the duality webs associated to those theories. These dualities where later understood as inherited (in the sense of \cite{Argyres:1999xu,Argyres:1999xu})
 from the $\mathcal{N}=2$ models of \cite{Uranga:1998uj} in \cite{Amariti:2021lhk}. In this paper we focus on a specific duality among the ones presented in \cite{Antinucci:2021edv,Amariti:2021lhk}, namely the duality that relates the two quiver theories above. Similarly to their $\mathcal{N}=2$ counterparts, these theories have a $\mathbb{Z}_2$ subgroup of the center that is not broken by the matter content and therefore can have different global structures. With an abuse of notation we use the same names for the $\mathcal{N}=1$ theories that we used in the $\mathcal{N}=2$ case, namely \eqref{eq:N2_SO_thys} and \eqref{eq:N2_USp_thys}. The duality orbits are shown in Figure \ref{eq:N2orbits}.\\
 
In this section we provide a check of these duality orbits by computing the Lens space index for small $n$ and small fugacities. In particular we perform the check with $n=2$ and $r=2,3$ and with $n=3$ and $r=2$. We find that the indices of theories in the same orbit are the same and the indices of theories in different orbits are different.

\subsection{$SO(5)\times SU(3)$ and $USp(2)\times SU(4)$}

\subsubsection*{r=2}
The contributions to the indices up to $\mathcal{O}(x^2)$ are:
\munutable{\mathcal{N}=1\quad\mathfrak{so}(5)\times\mathfrak{su}(3)}{2}
{3+9 x+21 x^2}
{2+6 x+4 x^{3/2}+18 x^2}
{1+x+5 x^2}
{2+4 x-4 x^{3/2}+8 x^2}

\munutable{\mathcal{N}=1 \quad \mathfrak{usp}(2)\times \mathfrak{su}(4)}{4}
{4+10 x+26 x^2}
{1+5 x+4 x^{3/2}+13 x^2}
{0}
{1+3 x-4 x^{3/2}+3 x^2}

The indices are:
\begin{equation}
	\mathcal{I}_{\mathcal{T}^{(Spin)}(4)} = \mathcal{I}_{\mathcal{T}^{(USp)}_+ (4)} = \mathcal{I}_{\mathcal{T}^{(USp)}_- (4)} = \mathcal{I}_{\mathcal{T}^{(SO)}_- (4)} = 3+9 x+21 x^2
\end{equation}
\begin{equation}
	\mathcal{I}_{\mathcal{T}^{(USp)}(4)} = \mathcal{I}_{\mathcal{T}^{(SO)}_+ (4)} =4+10 x+26 x^2
\end{equation}

\subsubsection*{r=4}
The contributions to the indices up to $\mathcal{O}(x^2)$ are:
\munutable{\mathcal{N}=1\quad\mathfrak{so}(5)\times\mathfrak{su}(3)}{4}
{8+29 x+12 x^{3/2}+70 x^2}
{6+26 x+16 x^{3/2}+64 x^2}
{4+7 x-4 x^{3/2}+16 x^2}
{2+4 x+10 x^2}

\munutable{\mathcal{N}=1\quad\mathfrak{usp}(2)\times\mathfrak{su}(4)}{4}
{10+33 x+12 x^{3/2}+80 x^2}
{4+22 x+16 x^{3/2}+54 x^2}
{2+3 x-4 x^{3/2}+6 x^2}
{0}

The indices are:
\begin{equation}
	\mathcal{I}_{\mathcal{T}^{(Spin)}(4)} = \mathcal{I}_{\mathcal{T}^{(USp)}_+ (4)} = \mathcal{I}_{\mathcal{T}^{(USp)}_- (4)} = \mathcal{I}_{\mathcal{T}^{(SO)}_- (4)} = 8+29 x+12 x^{3/2}+70 x^2
\end{equation}
\begin{equation}
	\mathcal{I}_{\mathcal{T}^{(USp)}(4)} = \mathcal{I}_{\mathcal{T}^{(SO)}_+ (4)} =10+33 x+12 x^{3/2}+80 x^2
\end{equation}

\subsection{$SO(7)\times SU(5)$ and $USp(4)\times SU(6)$}

\subsubsection*{r=2}
The contributions to the indices up to $\mathcal{O}(x^2)$ are:
\munutable{\mathcal{N}=1\quad\mathfrak{so}(7)\times\mathfrak{su}(5)}{2}
{4+11 x+31 x^2}
{3+10 x+4 x^{3/2}+27 x^2}
{2+7 x+15 x^2}
{3+8 x-4 x^{3/2}+19 x^2}

\munutable{\mathcal{N}=1\quad\mathfrak{usp}(4)\times\mathfrak{su}(6)}{2}
{6+18 x+46 x^2}
{1+3 x+4 x^{3/2}+12 x^2}
{0}
{1+x-4 x^{3/2}+4 x^2}

The indices are:
\begin{equation}
	\mathcal{I}_{\mathcal{T}^{(Spin)}(4)} = \mathcal{I}_{\mathcal{T}^{(USp)}_+ (4)} = \mathcal{I}_{\mathcal{T}^{(USp)}_- (4)} = \mathcal{I}_{\mathcal{T}^{(SO)}_- (4)} = 4+11 x+31 x^2
\end{equation}
\begin{equation}
	\mathcal{I}_{\mathcal{T}^{(USp)}(4)} = \mathcal{I}_{\mathcal{T}^{(SO)}_+ (4)} =6+18 x+46 x^2
\end{equation}

%
%
%
\section{A conformal triality}
\label{sec:triality}
In this section we consider the conformal triality introduced in \cite{Razamat:2020pra}. We compute the Lens space index for the three gauge theories involved. Each theory has a $\mathbb{Z}_2$ subgroup of the center that is not broken by the matter content and has three possible choices of global structure. We find that at small fugacities all nine theories have the same index, this suggests that they are all dual to each other. We argue that this is true by exploiting the fact that one of the frames describes the low energy theory of $D3$ branes on the tip of an orbifold $\mathbb{C}^3/{\mathbb{Z}_7}$. The holographic picture \cite{Bergman:2022otk} then suggests that the three choices of global structure for this frame are dual to each other. 
\subsection{Frame A}
The first frame is a $\mathcal{N}=1$ gauge theory with gauge algebra $\mathfrak{usp}(6)$ described by the quiver in \eqref{eq:frameA}.
\begin{equation}	\label{eq:frameA}
\includegraphics[scale=0.7]{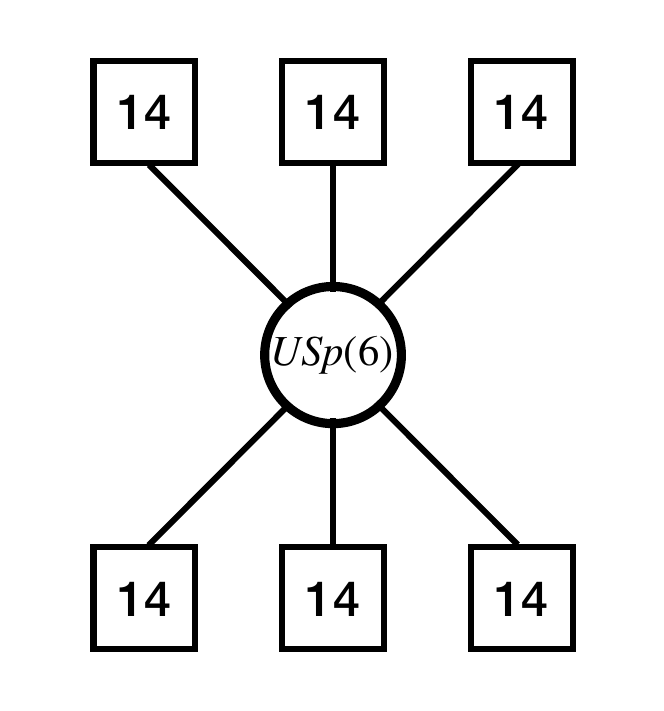}
\end{equation}
where 14 is the totally antisymmetric representation of $\mathfrak{usp}(6)$. The theory is conformal, all the matter fields have R-charge assignment $\frac{2}{3}$. With this assignment the one-loop beta function vanishes. The center of the gauge algebra $\mathbb{Z}_2$ is not broken by the matter content and the theory has three possible global structures:

\begin{equation}
\begin{split}
	\mathcal{T}^{(A)} = &\, USp(6)\\
	\mathcal{T}^{(A)}_+ =&\, \left( USp(6)/\mathbb{Z}_2 \right)_0 \\
	\mathcal{T}^{(A)}_- =&\, \left( USp(6)/\mathbb{Z}_2 \right)_1 \\
\end{split}
\end{equation}

The holonomy sectors can be organized by $(\mu,\nu) = (\pm1, \pm1)$ similarly to the cases studied in the previous sections. 

\subsubsection*{r=2}
The contributions from the $Z_{\mu,\nu}$ sectors up to $\mathcal{O}(x^2)$ are:

\munutable{\text{Frame A}}{2}
{2+42 x^{4/3}+42 x^2}
{1+21 x^{4/3}+42 x^2}
{0}
{1+21 x^{4/3}}

The indices are:
\begin{equation}
	\mathcal{I}_{\mathcal{T}^{(A)} } =\mathcal{I}_{\mathcal{T}^{(A)}_+ } =\mathcal{I}_{\mathcal{T}^{(A)}_- } = 2+42 x^{4/3}+42 x^2
\end{equation}

\subsubsection*{r=4}
The contributions from the $Z_{\mu,\nu}$ sectors up to $\mathcal{O}(x^2)$ are:

\munutable{\text{Frame A}}{4}
{3+63 x^{4/3}+84 x^2}
{2+42 x^{4/3}+84 x^2}
{1+21 x^{4/3}}
{0}

The indices are:
\begin{equation}
	\mathcal{I}_{\mathcal{T}^{(A)} } =\mathcal{I}_{\mathcal{T}^{(A)}_+ } =\mathcal{I}_{\mathcal{T}^{(A)}_- } = 3+63 x^{4/3}+84 x^2
\end{equation}

\subsection{Frame B}
The second frame is a $SU(2)^7$ gauge theory described by the quiver in \eqref{eq:frameB}.
\begin{equation}	\label{eq:frameB}
\includegraphics[scale=0.7]{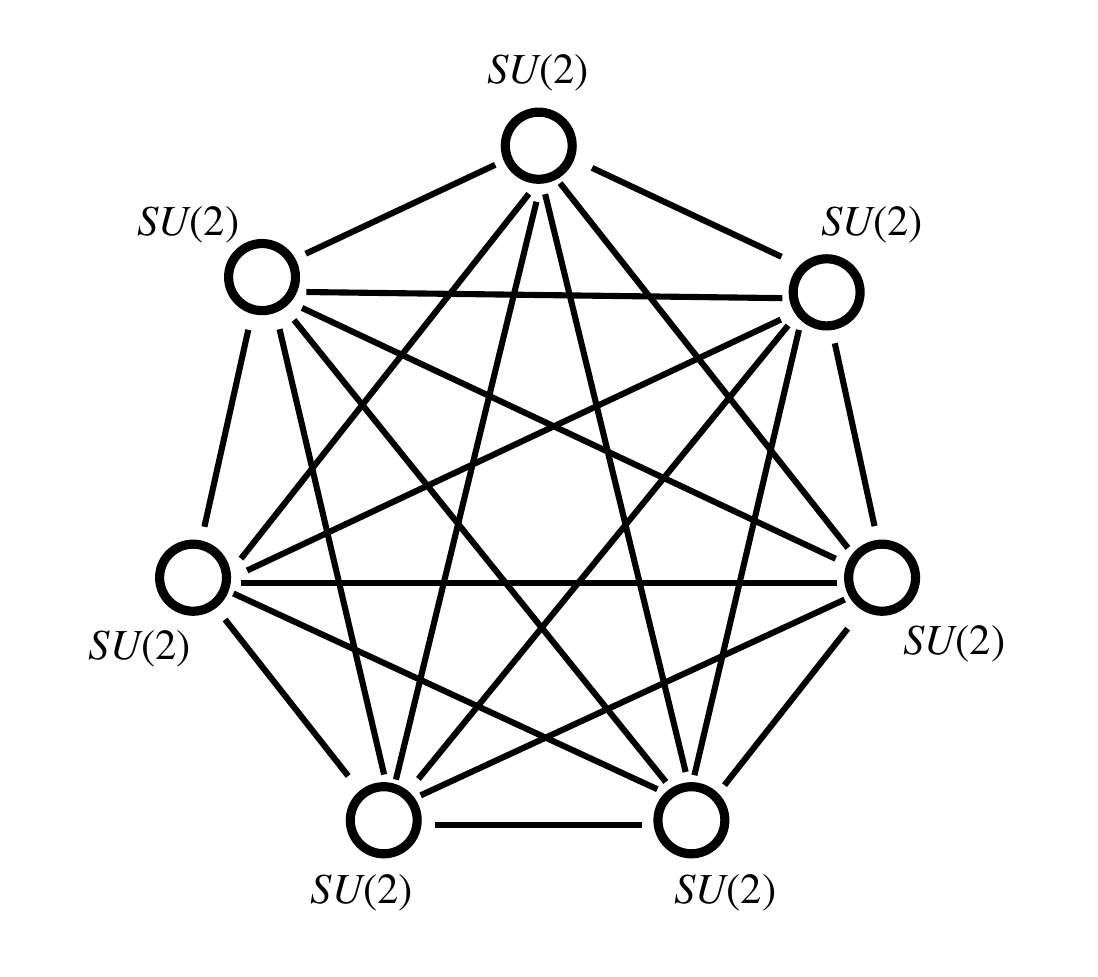}
\end{equation}
All the bifundamental matter fields have R-charge $\frac{2}{3}$. There is a diagonal $\mathbb{Z}_2$ subgroup of the center that is not broken by the matter content. There are three possible choices for the global structure:
\begin{equation}
\begin{split}
	\mathcal{T}^{(B)} = &\, SU(2)^7\\
	\mathcal{T}^{(B)}_+ =&\, \left( SU(2)^7/\mathbb{Z}_2 \right)_0 \\
	\mathcal{T}^{(B)}_- =&\, \left( SU(2)^7/\mathbb{Z}_2 \right)_1 \\
\end{split}
\end{equation}
The conformal manifold has complex dimension 21 \cite{Razamat:2020pra}. On a specific point of the conformal manifold the 
theory describes the low energy theory of two $D3$ branes in Type IIB probing the tip of the orbifold $\mathbb{C}^3/\mathbb{Z}_7^{(1,2,4)}$. 

\begin{figure}
\centering
	\includegraphics[scale=0.65]{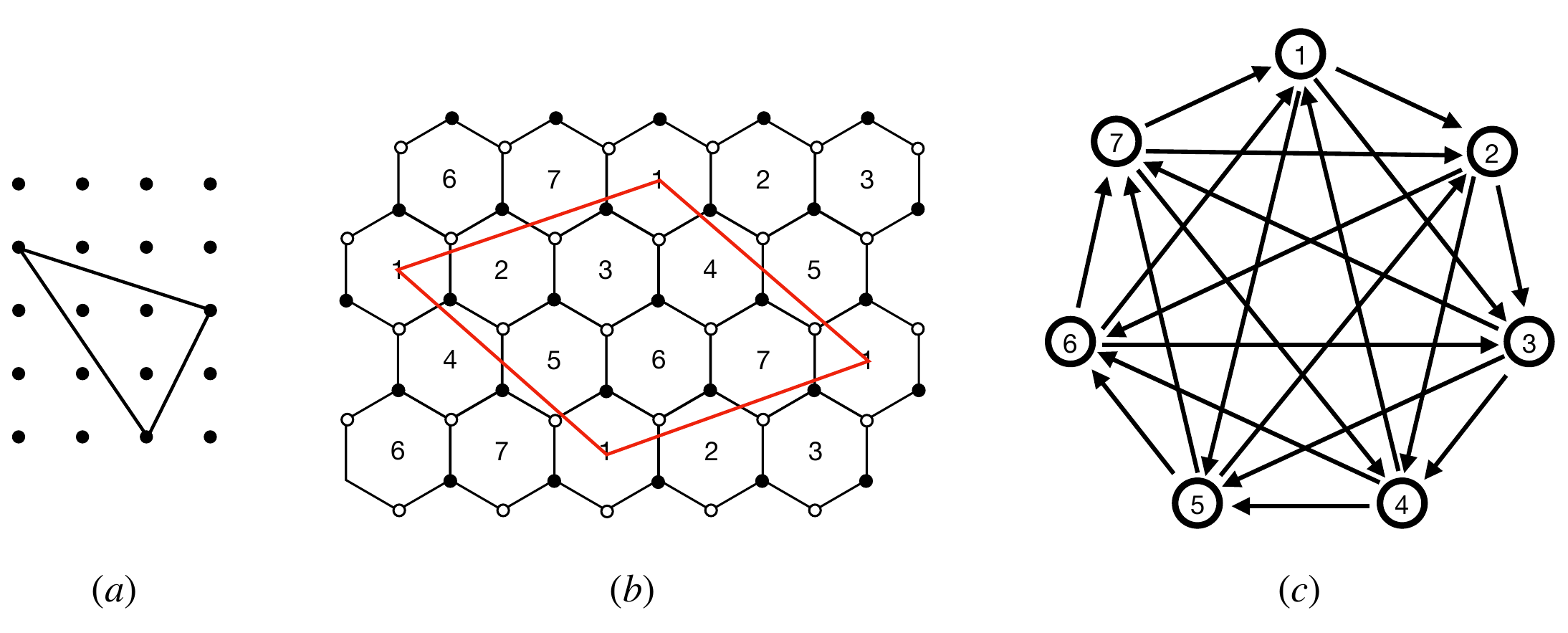}
	\caption{The toric diagram $(a)$, dimer model $(b)$ and quiver $(c)$ for the chiral orbifold $\mathbb{C}^3/\mathbb{Z}_7^{(1,2,4)}$. Each node in the quiver diagram corresponds to a $SU(n)$ gauge group, where $n$ is the number of $D3$ branes that prove the singularity.}
	\label{fig:C3Z7}
\end{figure}

In general the theory of $n$ $D3$ branes that probe this singularity is a $SU(n)^7$ toric theory described by  the dimer in Figure \ref{fig:C3Z7}. For general $n$ the quiver is oriented while for $n=2$ the bifundamental representations are pseudoreal and the theory is described by the unoriented quiver above.
 In the holographic description the choice of global structure is encoded in the boundary conditions for the Type IIB two-form fields $B_2$ and $C_2$ \cite{Aharony:1998qu,Witten:1998wy,Bergman:2022otk}. These are constrained by the five-dimensional CS action:
\begin{equation}
S_{C S}\left[B_2, C_2\right]=\int_{A d S_5 \times X_5} B_2 \wedge d C_2 \wedge d C_4=\frac{n}{2 \pi} \int_{A d S_5} B_2 \wedge d C_2
\end{equation}
where $X_5$ is the Sasaki-Einstein manifold whose real cone is $\mathbb{C}^3/\mathbb{Z}_7^{(1,2,4)}$. We notice that this is the same CS action as the one for $\mathcal{N}=4$ $SU(n)$ SYM, therefore the duality orbits of the $SU(2)^7$ theory will be mapped to the ones of $\mathcal{N}=4$ $SU(n)$ SYM. 
In the $n=2$ case that we are interested in the three choices of global structure for  the $\mathcal{N}=4$ theory are all dual, therefore we expect that $\mathcal{T}^{(B)}$, $\mathcal{T}^{(B)}_+$ and $\mathcal{T}^{(B)}_-$ lie in the same orbit.\\

\subsubsection*{r=2}
The contributions to the $Z_{\mu,\nu}$ sectors of the Lens space index up to $\mathcal{O}(x^2)$ are:
\munutable{\text{Frame B}}{2}
{2+42 x^{4/3}+42 x^2}
{1+21 x^{4/3}+42 x^2}
{1+21 x^{4/3}}
{0}

The indices are:
\begin{equation}
	\mathcal{I}_{\mathcal{T}^{(B)} } =\mathcal{I}_{\mathcal{T}^{(B)}_+ } =\mathcal{I}_{\mathcal{T}^{(B)}_- } = 2+42 x^{4/3}+42 x^2
\end{equation}

\subsubsection*{r=4}
The contributions to the $Z_{\mu,\nu}$ sectors of the Lens space index up to $\mathcal{O}(x^2)$ are:
\munutable{\text{Frame B}}{4}
{3+63 x^{4/3}+56 x^2}
{2+42 x^{4/3}+84 x^2}
{1+21 x^{4/3}}
{0}

The indices are:
\begin{equation}
	\mathcal{I}_{\mathcal{T}^{(B)} } =\mathcal{I}_{\mathcal{T}^{(B)}_+ } =\mathcal{I}_{\mathcal{T}^{(B)}_- } = 3+63 x^{4/3}+70 x^2
\end{equation}

\subsection{Frame C}
The second frame is a $SU(2)^2\times SU(4)$ gauge theory described by the quiver in \eqref{eq:frameC}.
\begin{equation}  \label{eq:frameC}
\includegraphics[scale=0.7]{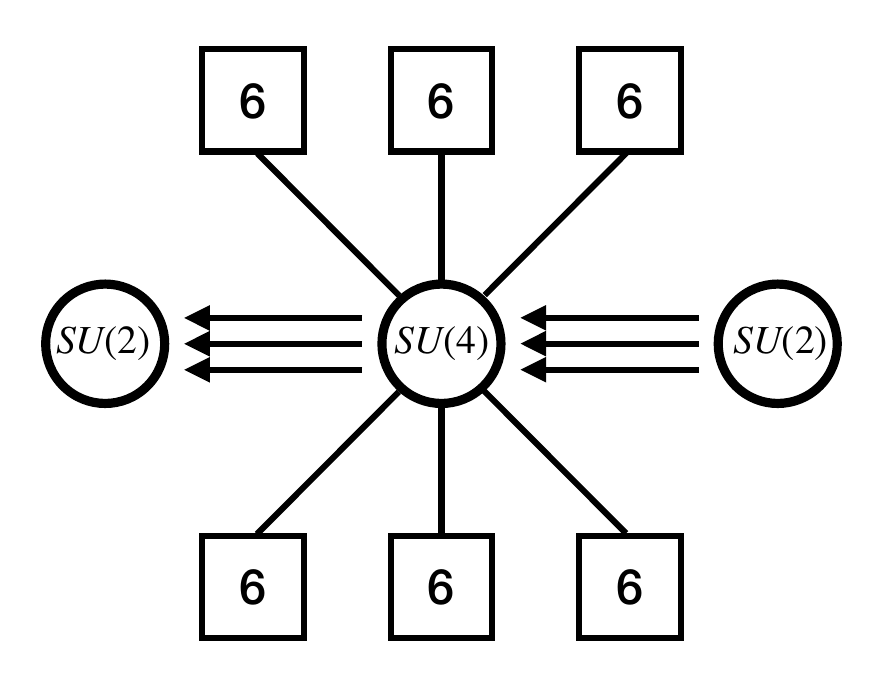}
\end{equation}
where 6 is the two-index antisymmetric representation of $\mathfrak{su}(4)$.
All the matter fields hare R-charge $\frac{2}{3}$. The theory has three possible global structures:
\begin{equation}
\begin{split}
	\mathcal{T}^{(C)} = &\, SU(2)\times SU(4) \times SU(2)\\
	\mathcal{T}^{(B)}_+ =&\, \left(\left( SU(2)\times SU(4) \times SU(2) \right)/\mathbb{Z}_2\right)_0 \\
	\mathcal{T}^{(B)}_- =&\, \left(\left( SU(2)\times SU(4) \times SU(2) \right)\right)_1 \\
\end{split}
\end{equation}
where $\mathbb{Z}_2$ is the diagonal subgroup of the center that is not broken by the matter content. 

\subsubsection*{r=2}
The contributions to the $Z_{\mu,\nu}$ sectors of the Lens space index up to $\mathcal{O}(x^2)$ are:
\munutable{\text{Frame B}}{2}
{2+42 x^{4/3}+42 x^2}
{1+21 x^{4/3}+42 x^2}
{1+21 x^{4/3}}
{0}

The indices are:
\begin{equation}
	\mathcal{I}_{\mathcal{T}^{(C)} } =\mathcal{I}_{\mathcal{T}^{(C)}_+ } =\mathcal{I}_{\mathcal{T}^{(C)}_- } = 2+42 x^{4/3}+42 x^2
\end{equation}

\subsubsection*{r=4}
The contributions to the $Z_{\mu,\nu}$ sectors of the Lens space index up to $\mathcal{O}(x^2)$ are:
\munutable{\text{Frame C}}{4}
{3+63 x^{4/3}+84 x^2}
{2+42 x^{4/3}+84 x^2}
{1+21 x^{4/3}}
{0}

The indices are:
\begin{equation}
	\mathcal{I}_{\mathcal{T}^{(C)} } =\mathcal{I}_{\mathcal{T}^{(C)}_+ } =\mathcal{I}_{\mathcal{T}^{(C)}_- } = 3+63 x^{4/3}+84 x^2
\end{equation}

The Lens space indices for the nine theories in the three frames match al low fugacities. This is a nontrivial check of the conformal triality conjectured in \cite{Razamat:2020pra}. Furthermore we propose that once the global properties of the three frames are taken into account all the nine theories that can be built lie on the same orbit. We argued that this can be understood from an holographic point of view in Frame B. It would be interesting to investigate this phenomenon in the other frames, A and C, we leave this discussion to future work.

\section{Conclusions}

In this paper we computed the Lens space for a series of models involving $USp(2n)$ gauge theories 
and various degrees of supersymmetry. We found the expected matching among models related by supersymmetric dualities and 
we also showed that the index does not coincide among different S-duality orbits.
In the analysis of the various models involving $USp(2n)$ gauge groups we encountered dual phases with orthogonal gauge groups
with odd rank.
While for the odd case the situation is quite under control in the even case the center is in general of order four and
more care is needed in the calculation of the almost commuting holonomies.
This problem does not emerge in the models studied in \cite{Razamat:2013opa} because the matter content breaks the center to an $\mathbb{Z}_2$ subgroup that exchanges the two spinorial representations.
The cases where the whole center or a different subgroup is preserved by the matter content has not be studied yet and 
it requires a separate analysis. 
The simplest examples corresponds to $SO(2n)$ $\mathcal{N}=4$ SYM. There are also interesting examples with lower supersymmetry, like the $\mathcal{N}=1$ S-dual models proposed in \cite{Etxebarria:2021lmq}.
Useful hits in this direction can be found in \cite{Imamura:2013qxa}.
Other $\mathcal{N}=4$ models that deserve an analysis are the one with exceptional gauge group and without a trivial center, i.e. the one with $E_{7}$ and $E_{8}$ algebra.
A last comment is related to the  $SU(n)^7$ orbifold that gives origin  the triality for $n=2$. This model for generic $n$ 
corresponds to a toric quiver gauge theory originating from a stack of D3 branes probing a Calabi-Yau toric threefold.
The lattice of charges for the line operators for models of this type (i.e. for toric quiver gauge theories) corresponds to the one obtained for $SU(n)$ $\mathcal{N}=4$ SYM. This can by shown by explicit analysis on the charge spectrum, as discussed in \cite{Amariti:2016hlj}.
It should be interesting to explain this behavior from a purely type IIB perspective, along the lines of \cite{Bergman:2022otk}.

 \section*{Acknowledgments}
This work has been supported in part by the Italian Ministero dell’Istruzione, Università e Ricerca (MIUR), in part by Istituto Nazionale di Fisica Nucleare (INFN) through the “Gauge Theories, Strings, Supergravity” (GSS) research project and in
part by MIUR-PRIN contract 2017CC72MK-003.

\pagebreak

\appendix
%
%
\section{Lens space index for symplectic gauge group}
\label{app:A}
%
%
\subsection{Almost commuting holonomies for $USp(2n)$}
The Lens space index of a supersymmetric gauge theory can be written as a sum/integral over the holonomies $g,h$ of the gauge field. These are classified by the solutions of:
\begin{equation} \label{app:commutator_g_h}
	g \cdot h \cdot g^{-1} \cdot h^{-1} = 1
\end{equation}

\begin{equation} \label{app:gr_eq1}
	g^r = 1
\end{equation}
modulo Weyl equivalence. In this section we are interested in computing the Lens space index for $USp(2n)$ and $USp(2n)/\mathbb{Z}_2$ gauge groups. We use the convention where the symplectic matrix is:
\begin{equation}
	\Omega_{2n\times 2n} = 
	\left(
		\begin{array}{cc}
			0_{n\times n} & -\text{I}_{n\times n} \\
			\text{I}_{n\times n} & 0_{n\times n}
		\end{array}
	\right)
\end{equation}
and a ${2n\times 2n}$ matrix $M$ is symplectic if:
\begin{equation}
	M^{T} \cdot \Omega_{2n\times 2n} \cdot M = \Omega_{2n\times 2n}.
\end{equation}

$USp(2n)$ is simply connected therefore $g$ and $h$ can be simultaneously conjugated to the maximal torus. They can be written as:
\begin{equation}	\label{app:g_USp}
	g = \text{diag} \left( e^{\frac{2 \pi i h_1}{r}},e^{\frac{2 \pi i h_2}{r}}, \dots, e^{\frac{2 \pi i h_n}{r}},
		e^{-\frac{2 \pi i h_1}{r}},e^{-\frac{2 \pi i h_2}{r}}, \dots, e^{-\frac{2 \pi i h_n}{r}}
		\right)
		\qquad 
		h_i \in \mathbb{Z}
\end{equation}
\begin{equation}	\label{app:h_USp}
	h = \text{diag} \left( w_1, w_2, \dots, w_n, \frac{1}{w_1},  \frac{1}{w_2}, \dots, \frac{1}{w_n}\right).
\end{equation}
The Weyl group acts as $h_i \leftrightarrow h_j$ and $h_i \to -h_i$, therefore we can take the $h_i$ such that:
\begin{equation}
	0 \leq h_1 \leq h_2 \leq \dots \leq h_n \leq \frac{r}{2}.
\end{equation}

In order to study the solutions to \eqref{app:commutator_g_h} and \eqref{app:gr_eq1} for $G=USp(2n)/\mathbb{Z}_2$ it is useful to uplift the holonomies to the universal cover $\tilde{G} = USp(2n)$:
\begin{equation} \label{app:commutator_g_h_nu}
	\tilde{g} \cdot \tilde{h} \cdot \tilde{g}^{-1} \cdot \tilde{h}^{-1} = \nu
\end{equation}
\begin{equation} \label{app:gr_mu}
	\tilde{g}^r = \mu
\end{equation}
where $\nu$ and $\mu$ are the possible uplifts of the element $1 \in G$. Generally they are element of the subgroup $H$ of the center of $\tilde{G}$ that we quotient by to obtain $G$. In this case $G=USp(2n)/\mathbb{Z}_2$ and $H = \mathcal{Z} (USp(2n)) = \lbrace \text{I}_{2n\times 2n},-\text{I}_{2n\times 2n} \rbrace$, therefore we can organize the solutions to  \eqref{app:commutator_g_h} and \eqref{app:gr_eq1} for $USp(2n)/\mathbb{Z}_2$ by the couple $(\nu,\mu) = (\pm 1, \pm 1)$. This is analogous to the case of orthogonal group studied in \cite{Razamat:2013opa}.\\

\paragraph{$(\nu,\mu) = (1,1)$:} The holonomies are the same as the ones for $USp(2n)$, \eqref{app:g_USp} and \eqref{app:h_USp}.

\paragraph{$(\nu,\mu) = (1,-1)$:} The holonomy $h$ is the same as the one for $USp(2n)$, \eqref{app:h_USp}, while:
\begin{equation}
	g = e^{\frac{\pi i}{r}}
		 \text{diag} \left( e^{\frac{2 \pi i h_1}{r}},e^{\frac{2 \pi i h_2}{r}}, \dots, e^{\frac{2 \pi i h_n}{r}},
		e^{-\frac{2 \pi i h_1}{r}},e^{-\frac{2 \pi i h_2}{r}}, \dots, e^{-\frac{2 \pi i h_n}{r}}
		\right)
		\qquad 
		h_i \in \mathbb{Z}.
\end{equation}
Due to Weyl equivalence we can take:
\begin{equation}
	0 \leq h_1 \leq h_2 \leq \dots \leq h_n < \frac{r}{2}.
\end{equation}

\paragraph{$(\nu,\mu) = (-1,1)$:} The solution to \eqref{app:commutator_g_h_nu} has been studied in \cite{Borel:1999bx}, here we only report the final result. For $USp(4k)$ we have:
\begin{equation}	\label{app:g_ac_4k}
	g = \text{diag} \left(
		\left[
			\begin{array}{cc}
			\lambda_1 & 0 \\
			0 & -\lambda_1
			\end{array}
		\right],
		\dots,
		\left[
			\begin{array}{cc}
			\lambda_k & 0 \\
			0 & -\lambda_k
			\end{array}
		\right],
		\left[
			\begin{array}{cc}
			\lambda_1^{-1} & 0 \\
			0 & -\lambda_1^{-1}
			\end{array}
		\right],
		\dots,
		\left[
			\begin{array}{cc}
			\lambda_k^{-1} & 0 \\
			0 & -\lambda_k^{-1}
			\end{array}
		\right]
	\right)
\end{equation}
\begin{equation}	\label{app:h_ac_4k}
	h = \text{diag}
		\left(
		\left[
			\begin{array}{cc}
			0 & w_1 \\
			w_1 & 0
			\end{array}
		\right],
		\dots,
		\left[
			\begin{array}{cc}
			0 & w_k \\
			w_k & 0
			\end{array}
		\right],
		\left[
			\begin{array}{cc}
			0 & w_1^{-1} \\
			w_1^{-1} & 0
			\end{array}
		\right],
		\dots,
		\left[
			\begin{array}{cc}
			0 & w_k^{-1}\\
			w_k^{-1} & 0
			\end{array}
		\right]
		\right).
\end{equation}
The additional constraint \eqref{app:gr_mu} with $\mu=1$ implies:
\begin{equation}
	\lambda_i = e^{\frac{2 \pi i h_i} {r}}
\end{equation}
and Weyl equivalence allows us to take:
\begin{equation}
	0 \leq h_1 \leq h_2 \leq \dots \leq h_k \leq \frac{r}{4}.
\end{equation}	

For $USp(4k+2)$ we have:
\begin{equation}	\label{app:g_ac_4k2}
	g = \text{diag} \left(
		\left[
			\begin{array}{cc}
			\lambda_1 & 0 \\
			0 & -\lambda_1
			\end{array}
		\right],
		\dots,
		\left[
			\begin{array}{cc}
			\lambda_k & 0 \\
			0 & -\lambda_k
			\end{array}
		\right],
		\left[
			\begin{array}{cc}
			i & 0 \\
			0 & -i
			\end{array}
		\right],
		\left[
			\begin{array}{cc}
			\lambda_1^{-1} & 0 \\
			0 & -\lambda_1^{-1}
			\end{array}
		\right],
		\dots,
		\left[
			\begin{array}{cc}
			\lambda_k^{-1} & 0 \\
			0 & -\lambda_k^{-1}
			\end{array}
		\right]
	\right)
\end{equation}
\begin{equation}	 \label{app:h_ac_4k2}
	h = \text{diag}
		\left(
		\left[
			\begin{array}{cc}
			0 & w_1 \\
			w_1 & 0
			\end{array}
		\right],
		\dots,
		\left[
			\begin{array}{cc}
			0 & w_k \\
			w_k & 0
			\end{array}
		\right],
		\left[
			\begin{array}{cc}
			0 & i \\
			i & 0
			\end{array}
		\right],
		\left[
			\begin{array}{cc}
			0 & w_1^{-1} \\
			w_1^{-1} & 0
			\end{array}
		\right],
		\dots,
		\left[
			\begin{array}{cc}
			0 & w_k^{-1}\\
			w_k^{-1} & 0
			\end{array}
		\right]
		\right).
\end{equation}

Equation \eqref{app:gr_mu} with $\mu=1$ only has solutions if $r = 0\, (\text{mod } 4)$, then we have:
\begin{equation}
	\lambda_i = e^{\frac{2 \pi i h_i} {r}}
\end{equation}
and Weyl equivalence allows us to take:
\begin{equation}
	0 \leq h_1 \leq h_2 \leq \dots \leq h_k \leq \frac{r}{4}.
\end{equation}	

\paragraph{$(\nu,\mu) = (-1,-1)$:} The solutions to \eqref{app:commutator_g_h_nu} are \eqref{app:g_ac_4k}, \eqref{app:h_ac_4k} and \eqref{app:g_ac_4k2}, \eqref{app:h_ac_4k2} for $USp(4k)$ and $USp(4k+2)$ respectively. For $USp(4k)$ the additional constraint \eqref{app:gr_mu} with $\mu=-1$ implies:
\begin{equation}
	\lambda_i = e^{\frac{\pi i}{r}} e^{\frac{2 \pi i h_i} {r}}
\end{equation}
with:
\begin{equation}
	0 \leq h_1 \leq h_2 \leq \dots \leq h_k < \frac{r}{4}.
\end{equation}	

For $USp(4k+2)$ the additional constraint \eqref{app:gr_mu} with $\mu=-1$ only has solutions if $r = 2\, (\text{mod } 4)$ where:
\begin{equation}
	\lambda_i = e^{\frac{\pi i}{r}} e^{\frac{2 \pi i h_i} {r}}
\end{equation}
with:
\begin{equation}
	0 \leq h_1 \leq h_2 \leq \dots \leq h_k < \frac{r}{4}.
\end{equation}	

We notice that the existence of solutions in the sectors with $\nu=-1$ depends on the $(\text{mod }4)$ behavior of $r$ and on the value of $\mu$. In particular we only have solutions when $r = 0\, (\text{mod } 4)$ and $\mu=1$ or when $r = 2\, (\text{mod } 4)$ and $\mu=-1$. This is a generalization of the mod $4$ behavior of $SU(2)$ almost commuting holonomies  already discussed in \cite{Razamat:2013opa}.\\

All things considered we find that the sectors with $\nu=1$ involve $n$ integrals and a sum over all the possible $h_i$ for a $USp(2n)$ theory, while the sectors with $\nu=-1$ involve $k$ integrals and a sum over the $h_i$ for $USp(4k)$ and $USp(4k+2)$ gauge theories. In the next sections we compute the contributions to the integrand given by the Haar measure, the two-index symmetric and (totally) antisymmetric representations and bifundamental fields between $USp(2n)$ and $SU(2m)$. We also  give the results for the contribution of the two-index symmetric and antisymmetric representations of $SU(2m)/\mathbb{Z}_2$ and the bifundamental representation of two unitary gauge groups.

\subsection{Symmetric and Antisymmetric representations for $USp(2n)$}
The almost commuting holonomies \eqref{app:g_ac_4k}, \eqref{app:h_ac_4k} for $USp(4k)/\mathbb{Z}_2$ (or \eqref{app:g_ac_4k2}, \eqref{app:h_ac_4k2} for $USp(4k+2)/\mathbb{Z}_2$) cannot be simultaneously diagonalized by conjugation of an element of the group. Their action on specific representations can however be diagonalized by choosing a proper basis for each representation (see \cite{Razamat:2013opa, Amariti:2019but} for some examples of this procedure). The two-index symmetric and antisymmetric representations of $USp(2n)$ can be represented by $2n\times 2n$ matrices, the natural basis for this space is $E^{(2n)}_{i,j}=e_i \times e_j$ where $e_i$, $i=1,\dots , n$ is a basis of the fundamental representation of $USp(2n)$. The holonomies act on a matrix $M$ in this space as:
\begin{equation}
	g \bullet M = g^T \cdot M \cdot g.
\end{equation}
In the sectors with $\nu=-1$ this action is not diagonal in the natural basis $E_{i,j}$. Therefore we define a new basis for this space:
\begin{equation}
	G^{(2n)}_{m;i,j} = H_m \times E^{(n)}_{i,j}
\end{equation}
where $H$ is the array of $2 \times 2$ matrices:
\begin{equation}
	H = \left\lbrace 
		\left(\begin{array}{cc}
1 & 0 \\
0 & 1
\end{array}\right),\left(\begin{array}{cc}
0 & 1 \\
1 & 0
\end{array}\right),\left(\begin{array}{cc}
-1 & 0 \\
0 & 1
\end{array}\right),\left(\begin{array}{cc}
0 & -1 \\
1 & 0
\end{array}\right)
	\right\rbrace.
\end{equation}
In this basis the action of the holonomies $g$ and $h$ is diagonal:
\begin{equation}
	g \bullet G^{(2n)}_{m;i,j} =  \eta^{(g)}_m e^{\frac{2 \pi i (\tilde{h}_i + \tilde{h}_j)}{r}} G^{(2n)}_{m;i,j} ,
	\qquad
	h \bullet G^{(2n)}_{m;i,j} = \eta^{(h)}_m \tilde{w}_i \tilde{w}_j G^{(2n)}_{m;i,j} 
\end{equation}
where:
\begin{equation}	\label{app:eta}
	\eta^{(g)}_m = \left\lbrace \begin{array}{ll}
				1 \qquad  &\text{if }  m=1,2 \\
				-1 \qquad &\text{if } m=3,4 
			\end{array},
			\right.
	\qquad
	\eta^{(h)}_m = \left\lbrace \begin{array}{ll}
				1 \qquad  &\text{if }  m=1,3 \\
				-1 \qquad &\text{if } m=2,4 
			\end{array},
			\right.
	.
\end{equation}
and:
\begin{equation} 	\label{app:htilde}
	\tilde{h} = \left\lbrace
	\begin{array}{ll}
		\lbrace h_i + \phi^{(\mu)} \rbrace_{i=1,\dots ,k} 
		\,\bigcup\,
		\lbrace - h_i - \phi^{(\mu)} \rbrace_{i=1,\dots ,k} 
		 &\quad\text{for } USp(4k) \\
		\lbrace h_i + \phi^{(\mu)} \rbrace_{i=1,\dots ,k} 
		\,\bigcup\,
		\lbrace \frac{r}{4} \rbrace
		\,\bigcup\,
		\lbrace - h_i - \phi^{(\mu)} \rbrace_{i=1,\dots ,k}  &\quad\text{for } USp(4k+2)
	\end{array}\right. 
\end{equation}
\begin{equation}	\label{app:wtilde}
	\tilde{w} = \left\lbrace
	\begin{array}{ll}
		( w_1, \dots, w_k, w_1^{-1}, \dots, w_k^{-1}) &\quad\text{for } USp(4k) \\
		( w_1, \dots, w_k, i,w_1^{-1}, \dots,w_k^{-1} ) &\quad\text{for } USp(4k+2)
	\end{array}\right.
\end{equation}
\begin{equation}
	\phi^{(\mu)} = \left\lbrace
		\begin{array}{ll}
			0 \qquad &\text{if } \mu=1 \\
			\frac{1}{2} \qquad &\text{if } \mu=-1
		\end{array}
	\right.
	.
\end{equation}

Notice that the eigenvalues for both $g$ and $h$ are symmetric in $i,j$, therefore we can define a basis for the symmetric and antisymmetric representations on which the action of $g$ and $h$ is diagonal as well. The basis for the symmetric representation is given by $G^{(2n)}_{m;i,j} + G^{(2n)}_{m;j,i}$ with $1 \leq i<j\leq n$ and $ m=1,2,3,4$ and by $G^{(2n)}_{m;i,i}$ for $i=1,\dots,n$ and $ m=1,2,3$. The basis for the antisymmetric representation is given by $G^{(2n)}_{m;i,j} - G^{(2n)}_{m;j,i}$ with $1 \leq i<j\leq n$ and $ m=1,2,3,4$ and by $G^{(2n)}_{4;i,i}$ for $i=1,\dots,n$. \\

The contribution of a chiral multiplet in the symmetric representation of $USp(2n)$ with R-charge $R$ in the  $(\mu, \nu) $ sectors is:
\begin{equation}	\label{app:S_11}
\begin{split}
\mathcal{S}^{(R)}_{1,1} (\mathbf{h},\mathbf{w}) =&
	\prod_{1 \leq i \leq j \leq n} \mathcal{I}_\chi^{(R)} \left( [\pm( h_i  + h_j) ], \,(w_i w_j)^{\pm 1} \right) \\
	&\prod_{1 \leq i < j \leq n} \mathcal{I}_\chi^{(R)} \left( [\pm( h_i  - h_j) ], \,(w_i / w_j)^{\pm 1} \right)
\end{split}
\end{equation}
\begin{equation} 	\label{app:S_-11}
\begin{split}
\mathcal{S}^{(R)}_{-1,1} (\mathbf{h},\mathbf{w})=&
	\prod_{1 \leq i \leq j \leq n} \mathcal{I}_\chi^{(R)} \left( [\pm( h_i  + h_j +1) ], \,(w_i w_j)^{\pm 1} \right) \\
	&\prod_{1 \leq i < j \leq n} \mathcal{I}_\chi^{(R)} \left( [\pm( h_i  - h_j) ], \,(w_i / w_j)^{\pm 1} \right)
\end{split}
\end{equation}
\begin{equation}	\label{app:S_mu-1}
\begin{split}
\mathcal{S}^{(R)}_{\mu,-1} (\mathbf{h},\mathbf{w}) =&
	\prod_{1 \leq i < j \leq \frac{n}{2}} \prod_{m=1}^4 
		\mathcal{I}_\chi^{(R)} \left([\tilde{h}_i + \tilde{h}_j + \phi^{(g)}_m ], \,\eta^{(h)}_m \tilde{w}_i \tilde{w}_j \right) \\
	& \prod_{1\leq i \leq \frac{n}{2}} \prod_{m=1}^3 
		\mathcal{I}_\chi^{(R)} \left([2 \tilde{h}_i + \phi^{(g)}_m ], \,\eta^{(h)}_m \tilde{w}_i^2  \right)
\end{split},
\end{equation}
where $\tilde{w}_i$ and $\tilde{h}_i$ where defined in eqs. \eqref{app:wtilde} and \eqref{app:htilde} respectively. $\phi^{(g)}_m$ is defined by $\eta^{(g)}_m = e^{2\pi i \phi^{(g)}_m /r}$. The contribution from the vector multiplet in the adjoint (symmetric) representation of $USp(2n)$ is given by \eqref{app:S_11}-\eqref{app:S_mu-1} with $\mathcal{I}_\chi^{(R)}$ replaced by $\mathcal{I}_V $.

The contribution from the antisymmetric representation of $USp(2n)$ is given by:
\begin{equation}	\label{app:A_11}
\begin{split}
\mathcal{A}^{(R)}_{1,1} (\mathbf{h},\mathbf{w}) =&
	\prod_{1 \leq i < j \leq n} \mathcal{I}_\chi^{(R)} \left( [\pm( h_i  + h_j) ], \,(w_i w_j)^{\pm 1} \right) \\
	&\prod_{1 \leq i < j \leq n} \mathcal{I}_\chi^{(R)} \left( [\pm( h_i  - h_j) ], \,(w_i / w_j)^{\pm 1} \right)
\end{split}
\end{equation}
\begin{equation} 	\label{app:A_-11}
\begin{split}
\mathcal{A}^{(R)}_{-1,1} (\mathbf{h},\mathbf{w}) =&
	\prod_{1 \leq i < j \leq n} \mathcal{I}_\chi^{(R)} \left( [\pm( h_i  + h_j +1) ], \,(w_i w_j)^{\pm 1} \right) \\
	&\prod_{1 \leq i < j \leq n} \mathcal{I}_\chi^{(R)} \left( [\pm( h_i  - h_j) ], \,(w_i / w_j)^{\pm 1} \right)
\end{split}
\end{equation}
\begin{equation}	\label{app:A_mu-1}
\begin{split}
\mathcal{A}^{(R)}_{\mu,-1} (\mathbf{h},\mathbf{w}) =&
	\prod_{1 \leq i < j \leq \frac{n}{2}} \prod_{m=1}^4 
		\mathcal{I}_\chi^{(R)} \left([\tilde{h}_i + \tilde{h}_j + \phi^{(g)}_m ], \,\eta^{(h)}_m \tilde{w}_i \tilde{w}_j \right) \\
	& \prod_{1\leq i \leq \frac{n}{2}} \prod_{m=4}^4
		\mathcal{I}_\chi^{(R)} \left([2 \tilde{h}_i + \phi^{(g)}_m ], \,\eta^{(h)}_m \tilde{w}_i^2  \right)
\end{split},
\end{equation}\\

Here $[m]$ is the number in $[0,r)$ such that $[m] = m \text{ mod }r$.
The Haar measure for $USp(2n)$ is:
\begin{equation}
\begin{split}
	\Delta_{1,1}^{(n)} (\mathbf{h},\mathbf{w})= &
		\frac{1}{|W_{1,1}^{(n)}(\mathbf{h})|}
		\prod_{1 \leq i \leq j \leq n} \left(\left( 1 - w_i w_j \right)
							\left( 1 - \frac{1}{w_i w_j} \right)\right)^{\delta_{ [h_i+h_j],0}}\\
		&\prod_{1 \leq i < j \leq n} \left(\left( 1 - \frac{w_i}{ w_j} \right)
							\left( 1 - \frac{w_j}{ w_i} \right)\right)^{\delta_{ [h_i-h_j],0}}
\end{split}
\end{equation}
\begin{equation}
\begin{split}
	\Delta_{-1,1}^{(n)} (\mathbf{h},\mathbf{w})= &
		\frac{1}{|W_{-1,1}^{(n)}(\mathbf{h})|}
		\prod_{1 \leq i \leq j \leq n} \left(\left( 1 - w_i w_j \right)
							\left( 1 - \frac{1}{w_i w_j} \right)\right)^{\delta_{ [h_i+h_j+1],0}}\\
		&\prod_{1 \leq i < j \leq n} \left(\left( 1 - \frac{w_i}{ w_j} \right)
							\left( 1 - \frac{w_j}{ w_i} \right)\right)^{\delta_{ [h_i-h_j],0}}
\end{split}
\end{equation}
\begin{equation}
\begin{split}
	\Delta_{\mu,-1}^{(n)}  (\mathbf{h},\mathbf{w})= &
		\frac{1}{|W_{\mu,-1}^{(n)}(\mathbf{h})|}
		\left. \prod_{1 \leq i < j \leq \frac{n}{2}} \prod_{m=1}^4 \right|_{\eta^{(h)}_m \tilde{w}_i \tilde{w}_j \neq 1}	
			\left(1 - \eta^{(h)}_m \tilde{w}_i \tilde{w}_j \right)^{\delta_{[\tilde{h}_i + \tilde{h}_j + \phi^{(g)}_m ],0}} \\
		&\left.  \prod_{1\leq i \leq \frac{n}{2}} \prod_{m=1}^3 \right|_{\eta^{(h)}_m \tilde{w}_i^2 \neq 1}
			\left(1-\eta^{(h)}_m \tilde{w}_i^2 \right)^{\delta_{[2 \tilde{h}_i + \phi^{(g)}_m ],0}} 
\end{split},
\end{equation}
where the factor $|W_{\mu,\nu}^{(n)}(\mathbf{h})|$ in the denominator ensures that the Haar measure is normalized:
\begin{equation}
	\oint \prod_{i=1}^n \frac{d w_i}{2 \pi i w_i} \Delta_{\mu,1}^{(n)}  (\mathbf{h},\mathbf{w})=1,
	\qquad
	\oint \prod_{i=1}^{\lfloor \frac{n}{2} \rfloor} \frac{d w_i}{2 \pi i w_i} \Delta_{\mu,-1}^{(n)}  (\mathbf{h},\mathbf{w})=1.
\end{equation}

\subsection{Symmetric and Antisymmetric representations for $SU(2n)$}
With a similar procedure to the one employed in the previous section we can compute the contributions of the two-index (conjugate) symmetric and (conjugate) antisymmetric representations of $SU(2n)$. We study $\mathcal{N}=1,2$ theories involving these representations in section \ref{sec:N2} and \ref{sec:N1}. The two-index representations leave a $\mathbb{Z}_2$ subgroup of the center $\mathcal{Z}(SU(2n))=\mathbb{Z}_{2n}$ unbroken, therefore they are representations of $SU(2n)/\mathbb{Z}_2$ as well.
Here we report the final result in the notation of \cite{Razamat:2013opa}:
\begin{equation}
\begin{aligned}	
\mathcal{S}^{(R)}_{1,1} (\mathbf m, \mathbf z) =&
	\prod_{1 \leq i \leq j \leq n} \mathcal{I}_\chi^{(R)} \left( [c( m_i  + m_j) ], \,(z_i z_j)^{c} \right) \\
%
\mathcal{S}^{(R)}_{-1,1} (\mathbf m, \mathbf z) =&
	\prod_{1 \leq i \leq j \leq n} \mathcal{I}_\chi^{(R)} \left( [c( m_i  + m_j) + 1 ], \,(z_i z_j)^{c} \right) 
\\
%
\mathcal{S}^{(R)}_{\mu,-1} (\mathbf m, \mathbf z) =&
	\prod_{1 \leq i < j \leq \frac{n}{2}} \prod_{m=1}^4 
		\mathcal{I}_\chi^{(R)} \left([c(m_i + m_j) + \phi^{(g)}_m ], \,\eta^{(h)}_m (z_i z_j)^c \right) \\
	& \prod_{1\leq i \leq \frac{n}{2}} \prod_{m=1}^3 
		\mathcal{I}_\chi^{(R)} \left([c(2 m_i) + \phi^{(g)}_m ], \,\eta^{(h)}_m z_i^{2c}  \right)
\\
%
\mathcal{A}^{(R)}_{1,1}(\mathbf m, \mathbf z)=&
	\prod_{1 \leq i < j \leq n} \mathcal{I}_\chi^{(R)} \left( [c( m_i  + m_j) ], \,(z_i z_j)^{c} \right) 
\\
%
\mathcal{A}^{(R)}_{-1,1} (\mathbf m, \mathbf z) =&
	\prod_{1 \leq i < j \leq n} \mathcal{I}_\chi^{(R)} \left( [c( m_i  + m_j) + 1 ], \,(z_i z_j)^{c} \right) 
\\
%
\mathcal{A}^{(R)}_{\mu,-1} (\mathbf m, \mathbf z) =&
	\prod_{1 \leq i < j \leq \frac{n}{2}} \prod_{m=1}^4 
		\mathcal{I}_\chi^{(R)} \left([c(m_i + m_j) + \phi^{(g)}_m ], \,\eta^{(h)}_m (z_i z_j)^c \right) \\
	& \prod_{1\leq i \leq \frac{n}{2}} \prod_{m=4}^4 
		\mathcal{I}_\chi^{(R)} \left([c(2 m_i) + \phi^{(g)}_m ], \,\eta^{(h)}_m z_i^{2c}  \right)
\end{aligned}
\end{equation}
where $c=1$ for the symmetric $\mathcal{S}$ (antisymmetric $\mathcal{A}$) representation and $c=-1$ for the corresponding conjugate representation.

\subsection{Bifundamental representations}
In this section we compute the contribution to the Lens space index integrand given by bifundamental fields. The case of a bifundamental field charged under two $SU(n)$ nodes was studied in \cite{Amariti:2019but}. We generalize this result to the case of a bifundamental field charged under two gauge groups $SU(n_1)$ and $SU(n_2)$ with different ranks and to the case of a bifundamental field charged a $USp(2n)$ group and a $SU(2m)$ group. The first case is used in the main body of the paper to compute the Lens space index of the theories studied in Section \ref{sec:triality} while the second is used in Sections \ref{sec:N2} and \ref{sec:N1}. At the end of this section we comment briefly on the contribution of bifundamental fields connecting an $SO(n)$ gauge group to another gauge group.\\

A bifundamental field between two gauge groups with algebras $\mathfrak{g}_1$ and $\mathfrak{g}_2$ breaks the center of the product algebra to a diagonal subgroup. In the cases we are interested in the center is broken as:
\begin{equation}
\begin{split}
\mathcal{Z} \left( \mathfrak{su}(n_1) \times \mathfrak{su}(n_2) \right) =&\, \mathbb{Z}_{n_1} \times \mathbb{Z}_{n_2} \quad\to\quad \mathbb{Z}_{ \text{gcd} (n_1,n_2)} \\
\mathcal{Z} \left( \mathfrak{usp}(2n) \times \mathfrak{su}(2m) \right) =&\, \mathbb{Z}_2 \times \mathbb{Z}_{2m} \quad\to\quad \mathbb{Z}_2
\end{split}
\end{equation}
therefore the bifundamental fields are consistent with the global forms:
\begin{equation}
\begin{gathered}
SU(n_1) \times SU(n_2),\\
\left( \left( SU(n_1) \times SU(n_2) \right) / \mathbb{Z}_k \right)_l
\end{gathered}
\end{equation}
with $k$ a divisor of $\text{gcd} (n_1,n_2)$ and
\begin{equation}
\begin{gathered}
USp(2n) \times SU(2m),\\
\left( \left( USp(2n) \times SU(2m) \right) / \mathbb{Z}_2 \right)_l
\end{gathered}
\end{equation}
respectively.\\

The flat connections for gauge fields with gauge algebra $\mathfrak{su}(n) \times \mathfrak{su}(m) $ are characterized by the holonomies around the non-contractible cycles of spacetime: $\gamma$ and the time cycle. They are respectively $g_1 \times g_2$ and $h_1 \times h_2$ where $g_1$ and $h_1$ are elements of the first group and $g_2$ and $h_2$ are elements of the second group. Their uplift to the universal cover groups $\tilde{g}_i$ and $\tilde{h}_i$ organize into sectors parametrized by $\mu, \nu$ that satisfy:
\begin{equation} \label{app:nu_bif_SU}
	\tilde{g}_1 \cdot \tilde{h}_1 \cdot \tilde{g}_1^{-1} \cdot \tilde{h}_1^{-1} = 
	\tilde{g}_2 \cdot \tilde{h}_2 \cdot \tilde{g}_2^{-1} \cdot \tilde{h}_2^{-1} =\nu=
	e^{\frac{2\pi i l_1}{n_1}}=
	e^{\frac{2\pi i l_2}{n_2}}
\end{equation}
\begin{equation} \label{app:mu_bif_SU}
	\tilde{g}_1^r = \tilde{g}_2^r =\mu =
	e^{\frac{2\pi i k_1}{n_1}} = e^{\frac{2\pi i k_2}{n_2}} 
\end{equation}
where $\mu$ and $\nu$ are elements of the subgroup of the center that we quotient by. The solutions for $g_i$ and $h_i$ are the same as the case of a single $SU(n)$ group \cite{Razamat:2013opa}. For further convenience we define:
\begin{equation}
d_i = \text{gcd}(n_i,l_i)\\
\end{equation}
and we notice that $\frac{d_1}{n_1}=\frac{d_2}{n_2}$.
 \\

A basis for the bifundamental representation is given by $E^{(n_1,n_2)}_{i,j} =e^{(n_2)*}_i\otimes e^{(n_1)}_i $ where $e^{(n_1)}_i$ is a basis of the fundamental representation of $\mathfrak{su}(n_1)$ and $e^{(n_2)*}_i $ is a basis of the antifundamental representation of $\mathfrak{su}(n_2)$. The holonomies act on a matrix  $B$ in this basis as:
\begin{equation}
\begin{split}
g \bullet B =& g_2^\dagger \cdot B \cdot g_1 \\
h \bullet B =& h_2^\dagger \cdot B \cdot h_1
\end{split}
\end{equation}
The action of the holonomies can be diagonalized by choosing a new basis for the bifundamental representation:
\begin{equation}
F^{(n_1,n_2)}_{p,q,i,j} = F^{(n_1/d_1)}_{p,q} \times E^{(d_1, d_2)}_{i,j}
\end{equation}
where:
\begin{equation}
	F^{(K)}_{p,q} = \sum_{j=1}^{K} e^{2\pi i p j / K} E^{(K,K)}_{j,K-j}
\end{equation}
This basis spans the space of the bifundamental representation with $p,q=1, \dots, n_1/d_1=n_2/d_2$ and $i=1,\dots , d_1$ and $j=1,\dots ,d_2$. The eigenvalues of the holonomies in this basis are:
\begin{equation}
	g \bullet F^{(n_1,n_2)}_{p,q,i,j} = e^{2\pi i q \frac{l_1}{n_1} +(m_i^{(1)} -m_j^{(2)}) } 
		F^{(n_1,n_2)}_{p,q,i,j},
	\qquad
	h\bullet F^{(n_1,n_2)}_{p,q,i,j} = e^{2\pi i p \frac{d_1}{n_1}} \frac{z_i^{(1)}}{z_j^{(2)}} 
		F^{(n_1,n_2)}_{p,q,i,j}
\end{equation}
the contribution to the Lens space integrand is:
\begin{equation}
B^{(n_1,n_2)}_{\mu,\nu} \left(\mathbf{m}^{(1)}, \mathbf{m}^{(2)},\mathbf{z}^{(1)},\mathbf{z}^{(2)}\right)= 
\prod_{p,q=0}^{n_1/d_1 - 1} \prod_{i=1}^{d_1}  \prod_{j=1}^{d_2}
\mathcal{I}_{\chi}^{(R)} \left( \left[m_i^{(1)} - m_j^{(2)}  + \frac{q l_1 r}{n_1} \right],
	e^{\frac{2\pi i p d_1}{n_1}} \frac{z_i^{(1)}}{z_j^{(2)}}
	\right)
\end{equation}
where $\mu, \nu$ are given in equations \eqref{app:nu_bif_SU}, \eqref{app:mu_bif_SU}. \\

The contribution from a field in the fundamental of $\mathfrak{usp}(2n)$ and the fundamental $\mathfrak{su}(2m)$ can be computed in a similar way. Here we report the result:
\begin{equation}
	C^{(2n,m)}_{\mu,1} \left( \mathbf{h}, \mathbf{m}, \mathbf{w}, \mathbf{z} \right) =
	\prod_{i=1}^{m} \prod_{j=1}^{2n}
	\mathcal{I}_{\chi}^{(R)} \left(
		\left[ h_i + m_j 	\right],
		w_i z_j
	\right)
	\mathcal{I}_{\chi}^{(R)} \left(
		\left[ - h_i + m_j \right],
		\frac{w_i}{z_j}
	\right)
\end{equation}

\begin{equation}
	C^{(2n,m)}_{\mu,-1} \left( \mathbf{h}, \mathbf{m}, \mathbf{w}, \mathbf{z} \right) =
	\prod_{i=1}^{m} \prod_{j=1}^{n}
	\prod_{k=1}^{4}
	\mathcal{I}_{\chi}^{(R)} \left(
	\left[ \tilde{h}_i + m_j +\phi_k^{(g)} \right],
	\tilde{w}_i z_j
	\right)
\end{equation}
similarly the contribution of a field in the fundamental of $\mathfrak{usp}(2n)$ and the antifundamental $\mathfrak{su}(2m)$ is obtained by substituting $z_i \to 1/z_i$ and $m_i \to -m_i$.
Here $\mathbf{h}, \mathbf{w}$ are associated to the holonomies for $\mathfrak{usp}(2n)$ and $\mathbf{m}, \mathbf{z}$ are associated to the holonomies for $\mathfrak{su}(2m)$. $\tilde{h}$, $\tilde{w}$ and $\phi_k^{(g)}$ are defined as in the case of a single symplectic group studied in the previous sections.

\subsection{Almost commuting holonomies for $SO(2n+1)$}
In this section we consider the almost commuting holonomies for $SO(2n+1)$ gauge theories that were studied in \cite{Razamat:2013opa}. We find a slightly different result than the one presented in the original paper, which is nevertheless crucial for matching the Lens space indices across the S-duality orbits of $\mathcal{N}=4$ SYM with orthogonal and symplectic gauge groups studied in Section \ref{sec:N4}.

The almost commuting holonomies for $SO(2n+1) = Spin(2n+1)/\mathbb{Z}_2$ are \cite{Razamat:2013opa}:
\begin{equation}
g=\left(\begin{array}{cccc}
-1 & 0 & 0 & 0 \\
0 & -1 & 0 & 0 \\
0 & 0 & 1 & 0 \\
0 & 0 & 0 & \mathcal{G}
\end{array}\right), \quad h=\left(\begin{array}{cccc}
-1 & 0 & 0 & 0 \\
0 & 1 & 0 & 0 \\
0 & 0 & -1 & 0 \\
0 & 0 & 0 & \mathcal{H}
\end{array}\right)
\end{equation}
where $\mathcal{G}$ and $\mathcal{H}$ are commuting matrices of $SO(2n-2)$. The Weyl symmetry of $SO(2n+1)$ include the Weyl symmetry of the subgroup $SO(2n-2)$ that act as $m_i \leftrightarrow m_j$ and $m_i \to -m_i$ for $i=1, \dots , n-2$ when applied to $g$\footnote{In this section we use the notation of \cite{Razamat:2013opa}.}. Using the notation of \cite{Razamat:2013opa} we can take:
\begin{equation} \label{app:m_razamat_willet}
\frac{r}{2} \geq m_1 \geq m_2 \geq \cdots \geq\left\| m_{n-1}\right\| \geq 0.
\end{equation}

The Weyl symmetry of $SO(2n+1)$ also include the transformation generated by the matrices:
\begin{equation}
w=\left(\begin{array}{cccc}
0 & 1 & 0 & 0 \\
1 & 0 & 0 & 0 \\
0 & 0 & 1 & 0 \\
0 & 0 & 0 & \mathcal{W}
\end{array}\right)
\end{equation}
where $\mathcal{W}$ is a $(2n-2)\times (2n-2)$ block-diagonal matrix with determinant $-1$ built out of the blocks $\left(\begin{array}{cc} 1 & 0 \\ 0 & 1 \end{array}\right)$ and $\left(\begin{array}{cc} 0 & 1 \\ 1 & 0 \end{array}\right)$. These matrices include the transformation $m_{n-1} \to -m_{n-1}$ that allows us to take:
\begin{equation}	\label{app:SO_m}
\frac{r}{2} \geq m_1 \geq m_2 \geq \cdots \geq m_{n-1}\geq 0.
\end{equation}
In the original paper \cite{Razamat:2013opa} equation \eqref{app:m_razamat_willet} was used to compute the possible values of $m_i$ for the $\nu=-1$ sector of $SO(2n+1)$ gauge theories. We argued that the correct values are given by equation \eqref{app:SO_m}. We have checked that in the computations performed in \cite{Razamat:2013opa} this does not make a difference, meaning that the holonomies are the same for the gauge groups considered and at low order in the Taylor expansion of the index. However by expanding the index at higher orders in the fugacities we find that the different values of $m_i$ given by the two formulae give different results for the index and we have checked that our result gives the correct match for theories that are believed to be dual, see section \ref{sec:N4} for example.

\bibliographystyle{JHEP}
\bibliography{ref}

\end{document}